\documentclass[tighten,nolinenumbers, notrackchanges, twocolumn]{aastex631}

\usepackage{amsmath}
\usepackage{amssymb}
\usepackage{amssymb}
\usepackage{mathtools}
\usepackage{mathrsfs}

\usepackage{morefloats}
\usepackage{longtable}
\usepackage{afterpage}

\hypersetup{linkcolor=red,citecolor=blue,urlcolor=magenta}

\usepackage{soul}

\usepackage[makeroom]{cancel}

\graphicspath{{figs/}}

\usepackage{xspace}

\newcommand*{\XPSI}{X-PSI\xspace}
\newcommand*{\NICER}{NICER\xspace}
\newcommand*{\xmm}{XMM-Newton\xspace}
\newcommand*{\PyMultiNest}{\textsc{PyMultiNest}\xspace}
\newcommand*{\MultiNest}{\textsc{MultiNest}\xspace}

\newcommand{\msol}{$M_\odot$\xspace}
\newcommand{\jdbl}{PSR~J0030$+$0451\xspace}
\newcommand{\joh}{PSR~J0740$+$6620\xspace}
\newcommand*{\jof}{PSR~J0437$-$4715\xspace}
\newcommand*{\joa}{PSR~J1231$-$1411\xspace}

\newcommand{\TT}[1]{\texttt{#1}}

\newcommand{\be}{\begin{equation}}
\newcommand{\ee}{\end{equation}}

\received{26 August 2024}
\revised{20 September 2024}
\accepted{21 September 2024}
\published{13 November 2024}

\submitjournal{The Astrophysical Journal}

\shorttitle{Radius of PSR J1231-1411}
\shorttitle{A NICER View of \joa}

\shortauthors{Salmi~et~al.}

\begin{document}

\title{A NICER View of \joa: A Complex Case}

\correspondingauthor{T.~Salmi}
\email{tuomo.salmi@helsinki.fi}
  
\author[0000-0001-6356-125X ]{Tuomo~Salmi}
\affil{Anton Pannekoek Institute for Astronomy, University of Amsterdam, Science Park 904, 1098XH Amsterdam, the Netherlands}
\affil{Department of Physics, University of Helsinki, P.O. Box 64, FI-00014, Finland}

\author[0000-0003-1226-0793]{Julia~S.~Deneva}
\affil{George Mason University, Resident at the U.S. Naval Research Laboratory, Washington, DC 20375, USA}

\author[0000-0002-5297-5278]{Paul~S.~Ray}
\affil{Space Science Division, U.S. Naval Research Laboratory, Washington, DC 20375, USA}

\author[0000-0002-1009-2354]{Anna~L.~Watts}
\affil{Anton Pannekoek Institute for Astronomy, University of Amsterdam, Science Park 904, 1098XH Amsterdam, the Netherlands}

\author[0000-0002-2651-5286]{Devarshi~Choudhury}
\affil{Anton Pannekoek Institute for Astronomy, University of Amsterdam, Science Park 904, 1098XH Amsterdam, the Netherlands}

\author[0000-0002-0428-8430]{Yves~Kini}
\affil{Anton Pannekoek Institute for Astronomy, University of Amsterdam, Science Park 904, 1098XH Amsterdam, the Netherlands}

\author[0000-0003-3068-6974]{Serena~Vinciguerra}
\affil{Anton Pannekoek Institute for Astronomy, University of Amsterdam, Science Park 904, 1098XH Amsterdam, the Netherlands}

\author[0000-0002-6039-692X]{H. Thankful Cromartie}
\affil{National Research Council Research Associate, National Academy of Sciences, Washington, DC 20001, USA}
\affil{Resident at Naval Research Laboratory, Washington, DC 20375, USA}

\author[0000-0002-4013-5650]{Michael~T.~Wolff}
\affil{Space Science Division, U.S. Naval Research Laboratory, Washington, DC 20375, USA}

\author[0009-0008-6187-8753]{Zaven~Arzoumanian}
\affil{X-Ray Astrophysics Laboratory, NASA Goddard Space Flight Center, Code 662, Greenbelt, MD 20771, USA}

\author[0000-0002-9870-2742]{Slavko~Bogdanov} 
\affil{Columbia Astrophysics Laboratory, Columbia University, 550 West 120th Street, New York, NY 10027, USA}

\author[0000-0001-7115-2819]{Keith~Gendreau}
\affil{X-Ray Astrophysics Laboratory, NASA Goddard Space Flight Center, Code 662, Greenbelt, MD 20771, USA}

\author[0000-0002-6449-106X]{Sebastien~Guillot}
\affil{Institut de Recherche en Astrophysique et Plan\'{e}tologie, UPS-OMP, CNRS, CNES, 9 avenue du Colonel Roche, BP 44346, F-31028 Toulouse Cedex 4, France}

\author[0000-0002-6089-6836]{Wynn~C.~G.~Ho}
\affil{Department of Physics and Astronomy, Haverford College, 370 Lancaster Avenue, Haverford, PA 19041, USA}

\author[0000-0003-4357-0575]{Sharon~M.~Morsink}
\affil{Department of Physics, University of Alberta, 4-183 CCIS, Edmonton, AB, T6G 2E1, Canada}

\author[0000-0002-1775-9692]{Isma\"{e}l Cognard}
\author[0000-0002-9049-8716]{Lucas Guillemot}
\affil{Laboratoire de Physique et Chimie de l'Environnement et de l'Espace, Universit\'e d’Orl\'eans/CNRS, 45071 Orl\'eans Cedex 02, France}
\affil{Observatoire Radioastronomique de Nan\c{c}ay, Observatoire de Paris, Universit\'e PSL, Universit\'e d’Orléans, CNRS, 18330 Nan\c{c}ay, France}

\author[0000-0002-3649-276X]{Gilles~Theureau}
\affil{Laboratoire de Physique et Chimie de l'Environnement et de l'Espace, Universit\'e d’Orl\'eans/CNRS, 45071 Orl\'eans Cedex 02, France}
\affil{Observatoire Radioastronomique de Nan\c{c}ay, Observatoire de Paris, Universit\'e PSL, Universit\'e d’Orléans, CNRS, 18330 Nan\c{c}ay, France}
\affiliation{Laboratoire Univers et Th\'eories LUTh, Observatoire de Paris, Universit\'e PSL, CNRS, Universit\'e de Paris, 92190 Meudon, France}

\author[0000-0002-0893-4073]{Matthew Kerr}
\affil{Space Science Division, U.S. Naval Research Laboratory, Washington, DC 20375, USA}

\begin{abstract}
Recent constraints on neutron star mass and radius have advanced our understanding of the equation of state (EOS) of cold dense matter.
Some of them have been obtained by modeling the pulses of three millisecond X-ray pulsars observed by the Neutron Star Interior Composition Explorer (NICER).
Here, we present a Bayesian parameter inference for a fourth pulsar, \joa, using the same technique with NICER and XMM-Newton data.
When applying a broad mass-inclination prior from radio timing measurements and the emission region geometry model that can best explain the data, we find likely converged results only when using a limited radius prior.
If limiting the radius to be consistent with the previous observational constraints and EOS analyses, we infer the radius to be $12.6 \pm 0.3$ km and the mass to be $1.04_{-0.03}^{+0.05}$ \msol,  each reported as the posterior credible interval bounded by the $16\,\%$ and $84\,\%$ quantiles.
If using an uninformative prior but limited between $10$ and $14$ km, we find otherwise similar results, but $R_{\mathrm{eq}} = 13.5_{-0.5}^{+0.3}$  km for the radius.
In both cases, we find a nonantipodal hot region geometry where one emitting spot is at the equator or slightly above, surrounded by a large colder region, and where a noncircular hot region lies close to southern rotational pole.
If using a wider radius prior, we only find solutions that fit the data significantly worse.
We discuss the challenges in finding the better fitting solutions, possibly related to the weak interpulse feature in the pulse profile.
\end{abstract}

\section{Introduction}
\label{sec:intro}

Modeling the X-ray pulses from a neutron star (NS) can be used to infer the NS's mass, radius, and the geometric properties of the X-ray emitting regions \citep[see e.g.,][and the references therein]{Pavlov1997,Watts2016,Bogdanov2019b}.
The inferred mass and radius can then be used to constrain the equation of state (EOS) of the supranuclear dense matter in NS cores \citep[see e.g.,][]{LP07}.
On the other hand, the emitting regions are associated with the NS's polar caps, and their properties can be used to probe the NS's magnetic field structure \citep[e.g.,][]{Bilous_2019,Carrasco2023}.

Previously, X-ray pulses of three NSs have been modeled using Neutron Star Interior Composition Explorer \citep[\NICER;][]{Gendreau2016} observations: \jdbl \citep{MLD_nicer19,Riley2019,Salmi2023,Vinciguerra2024bravo}, the massive pulsar \joh \citep{Miller2021, Riley2021, Salmi2022, Salmi2023, Dittmann2024, Salmi2024}, and the bright pulsar \jof \citep{Choudhury24}.
These have already provided useful constraints on dense matter models \citep[see, e.g.,][]{Miller2021, Raaijmakers2021, Biswas2022, Annala2023, Han2023, Takatsy23, Rutherford24}.
However, measuring masses and radii for a larger set of NSs and over a broad range of masses is expected to allow both tighter and more robust constraints.

\joa is a rotation-powered millisecond pulsar (RMP; discovered by \citealt{Ransom2011}), like many NSs previously analyzed with \NICER. 
Its thermally emitting surface regions are expected to be heated by the bombardment of charged particles from a magnetospheric return current \citep[see, e.g.,][]{RudermanSutherland1975,arons81,HM01}. 
Soft X-ray pulsations at the known spin frequency of $271$ Hz were found by \citet{Ray2019} and further analyzed in \citet{bogdanov19a}. 
The pulsed emission shows a broad and slightly asymmetric main pulse and a much fainter secondary interpulse (see Figure \ref{fig:data_hotel}). 
This feature differs from those seen in \NICER RMPs analyzed so far (see \citealt{bogdanov19a} for \jdbl and \jof, and \citealt{Wolff21} for \joh).
Similar to \joh and \jof, \joa is located in a binary system, allowing radio timing measurements to constrain the NS mass and observer inclination based on the relativistic Shapiro delay \citep{Fonseca20, Reardon24}. 
However, the obtained constraints in the preliminary radio timing results (available at the time of this work) are far less restrictive for \joa (Cromartie et al. 2024, in preparation) compared to \joh and \jof.
In this work, we use these measurements as prior distributions and explore the robustness of NS parameter inference for \joa.

The remainder of this paper is structured as follows. 
In Section \ref{sec:data_and_bkg}, we introduce the \NICER and \xmm data sets used for \joa.
In Section \ref{sec:methods}, we summarize the modeling procedure, and in Section \ref{sec:results} we present the results. 
We discuss the implications of the results in Section \ref{sec:discussion} and conclude in Section \ref{sec:conclusions}.
\setcounter{footnote}{14}
{
    \begin{figure}[t!]
    \centering
    \resizebox{\hsize}{!}{\includegraphics[
    width=\textwidth]{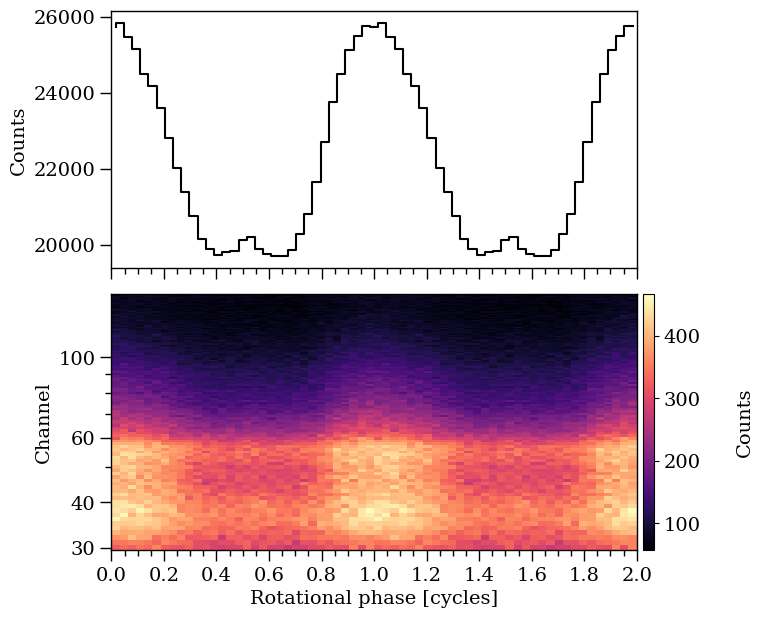}}
    \caption{\small{
The phase-folded \joa event data for two rotational cycles (for clarity).
The top panel shows the pulse profile summed over the channels. 
The total number of counts is given by the sum over all phase-channel pairs (over both cycles).
For the modeling, all the event data are grouped into a single rotational cycle instead, and thus each phase-channel bin has twice the number of counts shown here.
    }}
    \label{fig:data_hotel}
    \end{figure}
}






{
\begin{figure}[t!]
\centering
\resizebox{\hsize}{!}{\includegraphics[
width=\textwidth,scale=0.7]{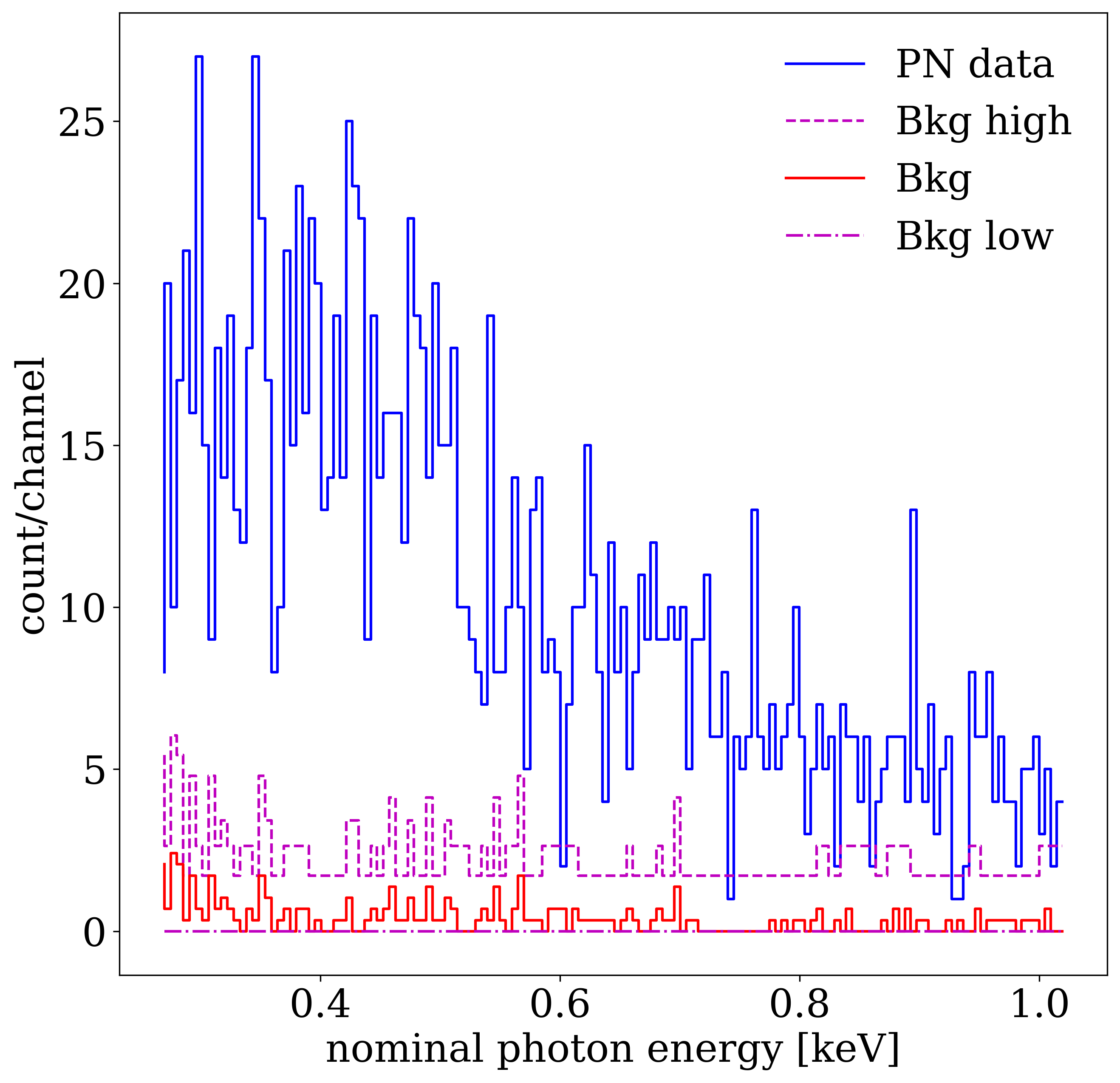}} 
\caption{\small{
\xmm EPIC-pn camera \joa count number spectrum as a function of nominal photon energies (blue-stepped curve). 
The measured rescaled background (red-stepped curve) and the background limits used in the analysis (purple dashed and dotted-dashed stepped curves) are also shown. 
The choice of the limits is explained in Section \ref{sec:likelihood_and_bkg}.
The complete figure set (3 images), including also the spectra and backgrounds for EPIC-MOS1 and EPIC-MOS2 instruments is available in the online journal (HTML version).
}}
\label{fig:data_xmm}
\end{figure}
}

\section{X-Ray Event Data}\label{sec:data_and_bkg}

\subsection{\NICER}\label{sec:data_nicer}

The NICER dataset for this work was produced using HEASoft 6.32.1\footnote{\url{https://heasarc.gsfc.nasa.gov/ftools}} \citep{heasoft2014} and CALDB xti20221001. We selected data from the full NICER mission up through 2023 May 14 (ObsID 6060060730), which is before NICER was affected by the light leak.\footnote{\url{https://heasarc.gsfc.nasa.gov/docs/nicer/analysis_threads/light-leak-overview/}}
After standard L2 processing with the task \texttt{nicerl2}, but before any of our filtering, the initial dataset contained about 3.53 Ms of exposure time. We applied further cuts to exclude data with magnetic cutoff rigidity $<1.5$ GeV$c^{-1}$; planetary K-index $K_p >5$; overshoot rate\footnote{\url{https://heasarc.gsfc.nasa.gov/docs/nicer/analysis_threads/overshoot-intro/}} $>1.5$ counts per second (cps) per Focal Plane Module (FPM); or a median undershoot rate\footnote{\url{https://heasarc.gsfc.nasa.gov/docs/nicer/analysis_threads/undershoot-intro/}} $>200$ cps/FPM. These cuts reduce the exposure to 81\% of the L2 exposure. Subsequently, we generated a background light curve using only 2--10 keV photons (where the pulsar flux is small so the count rate is background dominated) with 16 s bins and discarded times where that rate exceeded 1.25 cps. This removes the times polluted by high levels of trapped electron (TREL, in the nomenclature of the SCORPEON background model, see below) and low-energy electron (LEEL) backgrounds, reducing the exposure to 75\% of the L2 exposure. The final exposure was 2651393.0413 s.

The timing model to assign pulse phases to photon measured times was developed by first fitting a timing model to the radio time of arrivals from the Nan\c{c}ay radio telescope (NRT). Details on the NRT and on pulsar observations with this instrument can be found in \citet{Guillemot2023}. This determined the astrometric and binary parameters (including Shapiro delay) and a constant dispersion measure (DM). These were then held fixed while the frequency and frequency first derivative were fit to the Fermi-LAT gamma ray data, as done in  \citet{2022Sci...376..521F}. Finally, the reference epoch was set to
the values used in the Third Fermi-LAT Pulsar Catalog (3PC; \citealt{Smith2023}) so the absolute pulse phases will be consistent with that. This model provides pulse phases accurate to $<0.1\%$ of a pulse period over the entire duration of the NICER dataset.

The NICER science tool \texttt{nicerl3-spect} was used to compute a pulse height spectrum, response matrix, and background estimate using the SCORPEON background model.\footnote{\url{https://heasarc.gsfc.nasa.gov/docs/nicer/analysis_threads/scorpeon-overview/}} 
We used XSPEC \citep{Arn96} to fit the SCORPEON model parameters simultaneously with different source spectral models, including two blackbody components, one blackbody and one power-law component (with a power-law index fixed at $1.8$ as in \citealt{Ransom2011}), just a power-law component, one NS atmosphere component (\texttt{nsatmos}), and two NS atmosphere components.
These were all modified by interstellar absorption (e.g., \texttt{TBabs(BB + powerlaw)}).
For these data we found that, in addition to the standard free parameters in the SCORPEON model, we needed to free the normalization of the O\textsc{vii} line at 0.57 keV (presumably originating from solar wind charge exchange). 
In prior work before the development of SCORPEON \citep{Ray2019}, this line was added as an ad hoc Gaussian line.
We found that the obtained SCORPEON background estimates were notably dependent on the chosen source model, and therefore we predicted the background to range from the smallest estimate to the highest at each channel (after including also one standard deviation uncertainty for each model).
This range was used only in the end to compare our inferred background to it (see Section \ref{sec:results_pdtu}).

In the pulse profile analysis, we used the pulse invariant (PI) channel subset [30,150), corresponding to the nominal photon energy range [0.3, 1.5] keV, and 32 rotational phase bins.
The data split over two rotational cycles are visualized in Figure \ref{fig:data_hotel}.

\subsection{\xmm}\label{sec:data_xmm}

The XMM-Newton EPIC-MOS (\citealt{turner2001}; 1 and 2) and EPIC-pn \citep{struder2001} data (ObsID 0605470201) used in this study were obtained on 2009 July 15, as part of a Fermi/LAT unidentified source observation campaign \citep{Ransom2011}. 
We utilize the Scientific Analysis System software (SAS)\footnote{\url{https://www.cosmos.esa.int/web/xmm-newton/sas}} to analyze and build our data products, which in turn are input into the inference software. 
The fields of the new unidentified LAT sources were observed using the \xmm imaging mode, thus with insufficient time resolution to look for pulsations. 
Using the original raw data products from the EPIC-MOS and EPIC-pn observations, we extracted X-ray events from the now-known pulsar position.
We set the PATTERN $\leq$ 12 for the EPIC-MOS1 and EPIC-MOS2 instruments and PATTERN $\leq$ 4 for the EPIC-pn along with FLAG=0 for all three instruments. 
This resulted in 28.8 ks of exposure from the EPIC-MOS instruments and 17.7 ks of exposure for the EPIC-pn instrument.
For the EPIC-MOS1 and EPIC-MOS2 event data, we chose source+background extraction regions of $32\arcsec$ and source-free background extraction regions of $90\arcsec$. 
The extraction regions for source+background and background events for the EPIC-pn were $38\arcsec$ and $60\arcsec$, respectively. 
We filtered out times of higher-than-average particle noise from each event stream.
The resulting final list of events are stored in FITS files that are then converted into ASCII files containing just the event times and PI value.
However, no pulse phases are included in the XMM-Newton event files because the observations were made with a detector frame time that was too long to detect pulsations.
Once the event filtering is selected, we build X-ray spectra using the SAS tool ``evselect" and then use the SAS tools ``rmfgen" and ``arfgen" to build response matrices. 
The resulting response matrices are then turned into ASCII text files for ingesting into the inference code that is described below.
We also implicitly make the assumption that the average pulsar X-ray properties have not changed from the epoch of the XMM-Newton observation to the epoch of the NICER observations.
The assumption is justified since no evidence for long-term flux variability of \joa was found in \citet[][see their Section 7.3]{bogdanov19a}.

In the analysis, we used the channel subset [50,200) 
for the EPIC-pn instrument, and the channel subset [20, 100) 
for the both EPIC-MOS1 and EPIC-MOS2 instruments. 
These correspond to the nominal photon energy 
range [$0.27 - 1.02$ keV] for EPIC-pn and [$0.3 - 1.5$ keV] 
for EPIC-MOS1 and EPIC-MOS2.
The data are visualized in Figure \ref{fig:data_xmm}.

\section{Modeling Procedure}\label{sec:methods}

We use the X-ray Pulse Simulation and Inference\footnote{\url{https://github.com/xpsi-group/xpsi}} (\XPSI) code, with version \texttt{v2.1.1} for the inference runs and \texttt{v2.2.7} for producing the figures \citep{xpsi}.
Complete information of each run, data products, posterior sample files, and all of the analysis files can be found in the Zenodo repository: doi:\href{https:/doi.org/10.5281/zenodo.13358349}{10.5281/zenodo.13358349} \citep{salmi_zenodo24hotel}. 
In the next sections, we summarize the modeling procedure and focus on how it differs from that used in previous work.

\subsection{Effective-area Scaling Models}\label{sec:response}

As in previous works \citep{Riley2021,Choudhury24,Salmi2024,Vinciguerra2024bravo}, we capture the uncertainty in the effective area of both \NICER and all \xmm detectors by defining absolute energy-independent effective-area scaling factors.
Each factor consists of a shared and telescope specific component (see Section 2.2. in \citealt{Salmi2024}), and, as in most recent works, we select the uncertainties in those to be 10\% and 3\%, correspondingly, leading to a 10.4\% uncertainty in the full scaling factor.
As before, we commonly assume all scaling factors to be identical between EPIC-pn, EPIC-MOS1, and EPIC-MOS2 (leaving then only the scaling factors $\alpha_{\mathrm{NICER}}$ and $\alpha_{\mathrm{XMM}}$ as free parameters).
Since this may not be true, we have also performed a test run treating $\alpha_{\mathrm{XMM}}$ with $\alpha_{\mathrm{pn}}$, $\alpha_{\mathrm{MOS1}}$, and $\alpha_{\mathrm{MOS2}}$ as free independent parameters, although with all of them defined to have the same uncertainty (10.4\% as mentioned above) and the same correlation with $\alpha_{\mathrm{NICER}}$ (see Section \ref{sec:results_stu}).

\subsection{Pulse Profile Modeling Using X-PSI}\label{sec:pp_modeling}

As in previous \NICER analyses, we use the ``Oblate Schwarzschild'' approximation for the spacetime and NS shape model to calculate the energy-resolved X-ray pulses \citep[see e.g.,][]{ML98,NS02,PG03,CMLC07,MLC07,lomiller13,AGM14,Bogdanov2019b,Watts2019}. 
The spectrum and the beaming pattern of the surface radiation are in most cases assumed to follow those from a fully ionized hydrogen atmosphere model \texttt{NSX} \citep{Ho01}, but we also performed a run assuming fully ionized helium \texttt{NSX} atmosphere (see Section \ref{sec:results_pdtu}).
In addition, for one case we also modeled the NS surface outside the hot regions, assuming it emits blackbody radiation (for computational efficiency).
This assumption is more realistic than having no surface emission, although more accurate still would be to use a numerical (and thus more expensive) partially ionized atmosphere model for the likely relatively cold surface.
To account for the interstellar attenuation, we use the \texttt{TBabs} model \citep[][updated in 2016]{Wilms2000}.

To characterize the geometry (and number) of the hot regions, we apply two different models: Single Temperature-Unshared (\texttt{ST-U}) and Protruding Dual Temperature-Unshared (\texttt{PDT-U}).
In the former model two hot regions are described by two single-temperature disjoint spherical caps with unshared parameters. 
The latter model is the most complex model applied with \XPSI so far, where each of the two hot regions is additionally described by two overlapping spherical caps with different temperatures (see \citealt{Riley2019} and Figure 1 of \citealt{vinciguerra2023sim} for more details).
Where the two overlap, emission of only one of them is accounted for, and that component is referred to as being ``superseding" (the other being ``ceding").

For the pulse profile modeling resolution with \XPSI, we mostly follow the settings\footnote{For more details see Section 2.3.1 in \citet{Vinciguerra2024bravo} or \XPSI documentation at \url{https://xpsi-group.github.io/xpsi/hotregion.html}.} used in the previous \joh analyses \citep{Riley2021,Salmi2022,Salmi2024}. 
For example, we calculate multiple images (photons reaching the observer from the same point at the NS surface with many deflection angles due to high gravity) up to third order in case they are visible.
The initial deflection angle grid is again calculated for 512 different emission angles at each colatitude.
However, to speed-up the calculation we reduced the number of model energies (from 128 to 64) and phases (from 64 to 32) for the more expensive \texttt{PDT-U} model.
This change was tested and found to have no effects on the results for a \texttt{ST-U} run.
In contrast, the resolution of the hot region cell mesh (\texttt{sqrt\_num\_cells} = 32, \texttt{min\_sqrt\_num\_cells} = 10, \texttt{max\_sqrt\_num\_cells} = 64) was not reduced to avoid loss of accuracy in the computed waveforms for extreme configurations, as found by \citet{Choudhury24raytracing}.

\subsection{Priors}\label{sec:priors}

\subsubsection{Mass and Inclination Priors}\label{sec:priors_mass_inclination}

We use mass $M$ and inclination $i$ priors that are based on the preliminary Shapiro delay measurement results from Cromartie et al. (2024, in preparation), which were obtained through a Bayesian timing model fit to data from the Robert C. Byrd Green Bank Telescope and from the NRT. 
To approximate the 2D probability density function, we first fitted a Gaussian function to the marginalized probability distribution of $M$ using a subset of the Markov chain Monte Carlo samples from the Shapiro delay analysis (those having $M$ between 1 and 3 \msol).
We obtained a mean of $\sim 1.002$ \msol and a standard deviation of $\sim 0.930$ \msol.
We inverse sampled the mass from this distribution (truncating it between 1 and 3 \msol). \footnote{We note that the resulting prior is still quite close to uniform.}
After that, we inverse sampled the $\cos i$ from another Gaussian, whose mean and standard deviations were set to depend on the value of the already drawn mass; $\mu_{\cos i}(M)$, $\sigma_{\cos i}(M)$. 
The values were found by fitting Gaussians to the $\cos i$ probability distributions for a grid of masses and by fitting their dependence on mass with a polynomial function.
This resulted in the following:
\begin{equation}\label{eq:mu_cosi}
\mu_{\cos i}(M) = a_{\mu} \Big(\frac{M}{M_\odot}\Big)^{2}+ b_{\mu} \frac{M}{M_\odot} + c_{\mu}
\end{equation}
and
\begin{equation}\label{eq:sigma_cosi}
\sigma_{\cos i}(M) = a_{\sigma} \Big(\frac{M}{M_\odot}\Big)^{2}+ b_{\sigma} \frac{M}{M_\odot} + c_{\sigma},
\end{equation}
where $a_{\mu} = -0.00835942$, $b_{\mu} = 0.10890304$, $c_{\mu} = 0.01118702$, $a_{\sigma} = -0.00097777$, $b_{\sigma} = 0.01013241$, and $c_{\sigma} = 0.01509739$.

The resulting 2D prior for $M$ and $\cos i$ is shown in Figure \ref{fig:mcosi_prior}.
We see that highest probabilities are at the smallest allowed masses but that the distribution fades very slowly toward higher masses.
Inclination is slightly better constrained, peaking above $80\degr$ ($\cos i \lesssim 0.18$).
We note that the final prior adopted in \XPSI is additionally modified by the compactness and surface gravity conditions (as e.g., in \citealt{Riley2021}) to prevent the polar radius from being smaller than the photon sphere ($R_{\textrm{polar}}/r_{\rm g}(M)>3$, where $r_{\rm g}$ is the Schwarzchild radius), and to not extrapolate intensities beyond the precomputed atmosphere grid (e.g., $13.7\leq \log_{10}g\leq15.0$, for the fully ionized hydrogen grid, where $g$ is the surface gravity).
This makes the prior probability at smaller masses even higher because fewer NS samples fail the compactness condition there, hence more are retained.

{
\begin{figure}[t!]
\centering
\resizebox{\hsize}{!}{
\includegraphics[
width=\textwidth]{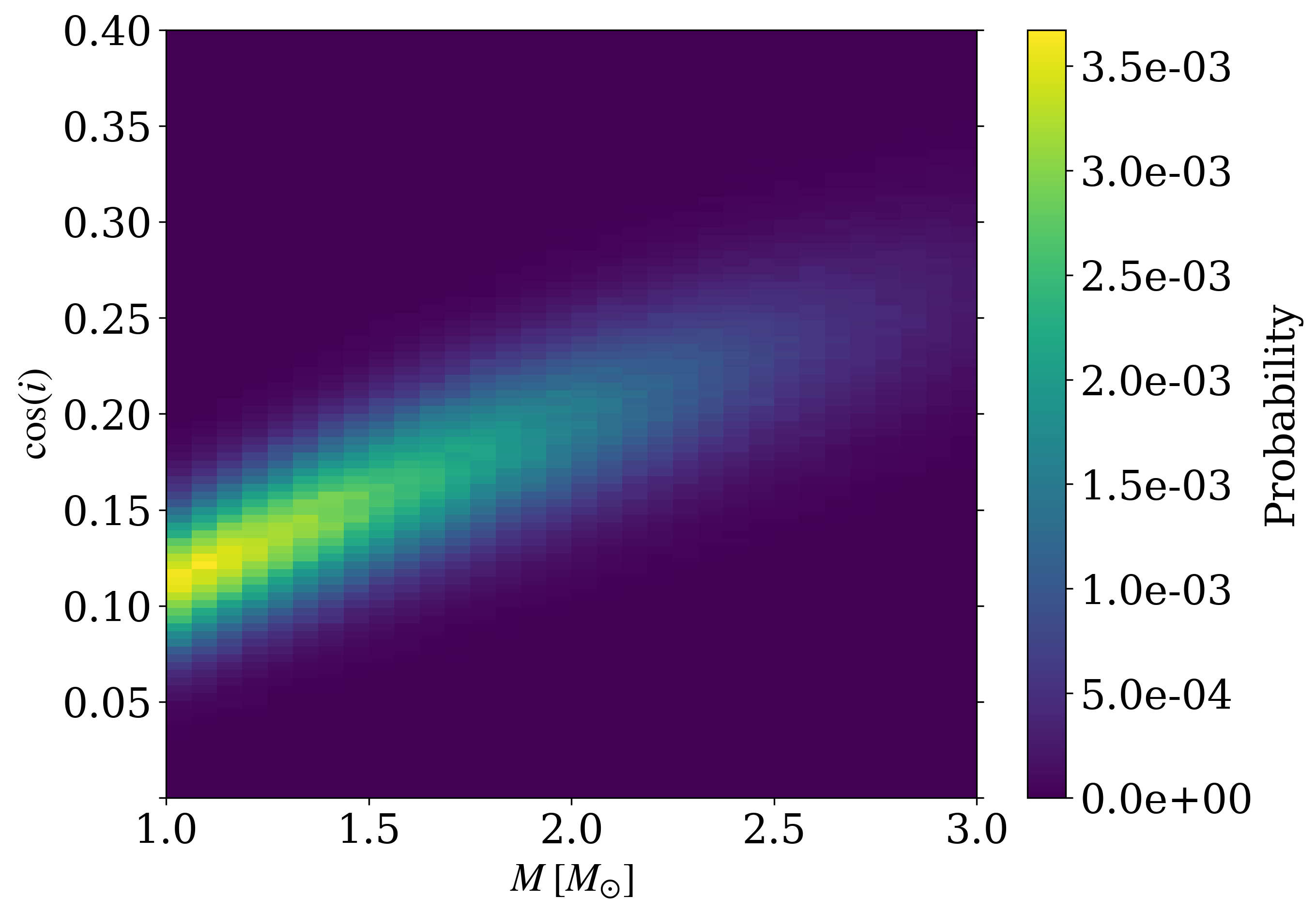}
}    
\caption{\small{
A 2D $\cos i - M$ prior distribution based on the radio timing measurements from Cromartie et al. (2024, in preparation). 
The color indicates the probability of a $\cos i - M$ cell.
}}
\label{fig:mcosi_prior}
\end{figure}
}

\subsubsection{Radius Prior}\label{sec:radius_prior}

For the NS (equatorial) radius, \XPSI analyses have usually assumed a prior that is initially flat between $\sim 4.4$ km (i.e., $3r_{\rm g}$ for a $1$ \msol star) and $16$ km, but then modified by the compactness and surface gravity conditions, mentioned also in Section \ref{sec:priors_mass_inclination}. 
In this work, we apply also more informative radius priors. 
In particular, we consider four cases when choosing a radius prior.
In the first scenario, we use the standard \XPSI radius prior.
In the second scenario, we set the lower limit of radius to be $8$ km to avoid extremely compact solutions.
In the third scenario, we restrict the radius to be between $10$ and $14$ km.
In the fourth scenario (called ``R21"), we require the radius to be consistent with previous observational and theoretical EOS constraints by using the posterior samples from \citet[][on Zenodo, \citealt{Raaijmakers2021_zenodo}]{Raaijmakers2021} to numerically build up a 1D prior distribution for radius.
We selected the samples from the ``combined constraints" case obtained with the ``PP" parametrization of the EOS corresponding roughly a Gaussian centered around 12 km (see the lower left-hand panel in Figure 5 of \citealt{Raaijmakers2021}). 
We converted the samples to a cumulative probability distribution using the \texttt{numpy.cumsum} function and interpolated from that distribution with \texttt{scipy.interpolate.Akima1DInterpolator} when sampling.

\subsubsection{Other Priors}\label{sec:other_priors}

For all the other model parameters, we use fairly uninformative priors.
For the geometry parameters describing the hot regions, we follow the choices of \citet{Riley2021} and \citet{Vinciguerra2024bravo}, i.e., we have initially uniform priors in the spherical cap angular radii (limited to be below $\pi/2$), in their phase coordinates and in cosines of their colatitudes. 
These priors are then modified by the requirements that primary and secondary hot regions cannot overlap (but in case of \texttt{PDT-U} the two caps forming a hot region must overlap), and that the primary hot region is always the one with smaller colatitude.
The hot region temperatures are uniform in $\log_{10}T_{\mathrm{eff}}$ and bound based on the limits of the precomputed atmosphere model grid (e.g., $\log_{10}T_{\mathrm{eff}} \in [5.1, 6.8]$ for fully ionized hydrogen).
When including blackbody emission from the rest of the star's surface, we sample its temperature $\log_{10}{T_{\mathrm{else}}}$ uniformly between $5.0$ and $6.5$ as in \citet{Vinciguerra2024bravo}.

Unlike for the previous NSs analyzed with \NICER, the distance $D$ to \joa is known rather poorly.
\citet{Yao2017} estimated the $D$ to be $420$ pc using the DM of \joa and an electron density model of the Galaxy.
However, they also reported a distance range between $350$ and $510$ pc based on independent estimates from optical measurements of the white dwarf companion. 
In this work, we use slightly more conservative limits and sample $D$ uniformly between $100$ and $700$ pc.
We note, though, that preliminary parallax distance measurements indicate a distance of roughly $600 \pm 100$ pc (Cromartie et al. 2024, in preparation).

Since the interstellar hydrogen column density $N_{\mathrm{H}}$ is also not well known, we sample that uniformly between $0$ and $10^{21} \mathrm{cm}^{-2}$, as in \citet{Riley2021}.
In addition, the effective-area scaling factor priors are those described already in Section \ref{sec:response}.

\subsection{Likelihood and Background}\label{sec:likelihood_and_bkg}

We construct a joint \NICER and \xmm likelihood function and marginalize the likelihoods over background variables (one for each detector channel) in the same way as described in Section 2.5 of \citet{Riley2021}.
We do not directly impose any informative limits on the \NICER background, but for \xmm instruments we bound the background marginalization to be done within $\mathscr{B}_{\rm X} \pm 4 \sqrt{\mathscr{B}_{\rm X}}$ (but always above $0$), where $\mathscr{B}_{\rm X}$ is the measured number of background counts for a given channel (see the dashed lines in Figure \ref{fig:data_xmm} after being rescaled based on the BACKSCAL factor, i.e., the difference in the extraction area between the \joa and background observation).
The prior in the background is assumed to be uniform between the lower and upper limits.
Constraining the background using \xmm also offers an indirect constraint on the \NICER background by restricting the number of source photons that can satisfy both \NICER and \xmm data simultaneously.
In the end, we cross-check the inferred \NICER background with that predicted by the SCORPEON background model for \NICER (see Sections \ref{sec:data_nicer} and \ref{sec:results_pdtu}). 

\subsection{Posterior Computation}\label{sec:posterior_computation}

We compute the posterior samples using \PyMultiNest \citep{PyMultiNest} and \MultiNest \citep{MultiNest_2008,multinest09,FHCP2019}, as in the previous \XPSI analyses.
In all runs, we set both the sampling efficiency (defined as described in Appendix B of \citealt{Salmi2024}) and the evidence tolerance to $0.1$.
We also keep the multimode/mode-separation setting of \MultiNest turned on to get more information about the multiple modes found in the posterior volume of \joa. 
We checked for one \texttt{ST-U} test run that turning the multimode setting off had no significant effects on the results or on the computation time.
For \texttt{ST-U} runs, we used 4000 live points, which was deemed sufficient based on preliminary test runs with older data and mass-inclination prior versions.
For \texttt{PDT-U} runs, we increased the number of live points to 10,000 and to 20,000.
However, these were found to provide likely converged results only in the case of the two most restricted radius priors (the third and fourth cases described in Section \ref{sec:radius_prior}).
Due to the already high computational cost (above 200,000 core hours for a 20,000 live point run), better convergence by increasing the live points further or by dropping the sampling efficiency was not attempted.

{
    \begin{figure*}[!htbp]
    \centering
    \includegraphics[
    width=0.49\textwidth]
    {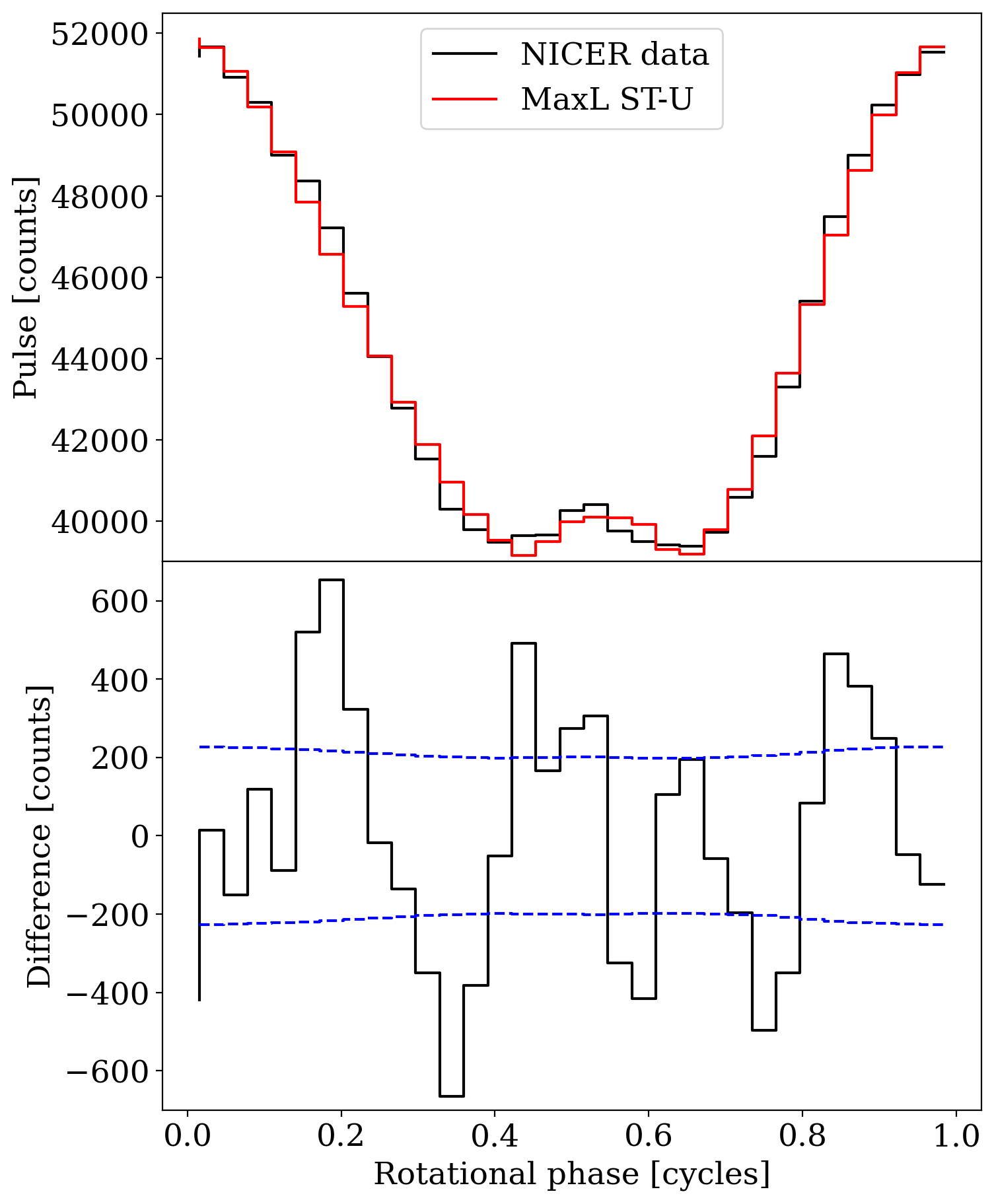}
    \includegraphics[
    width=0.49\textwidth]
    {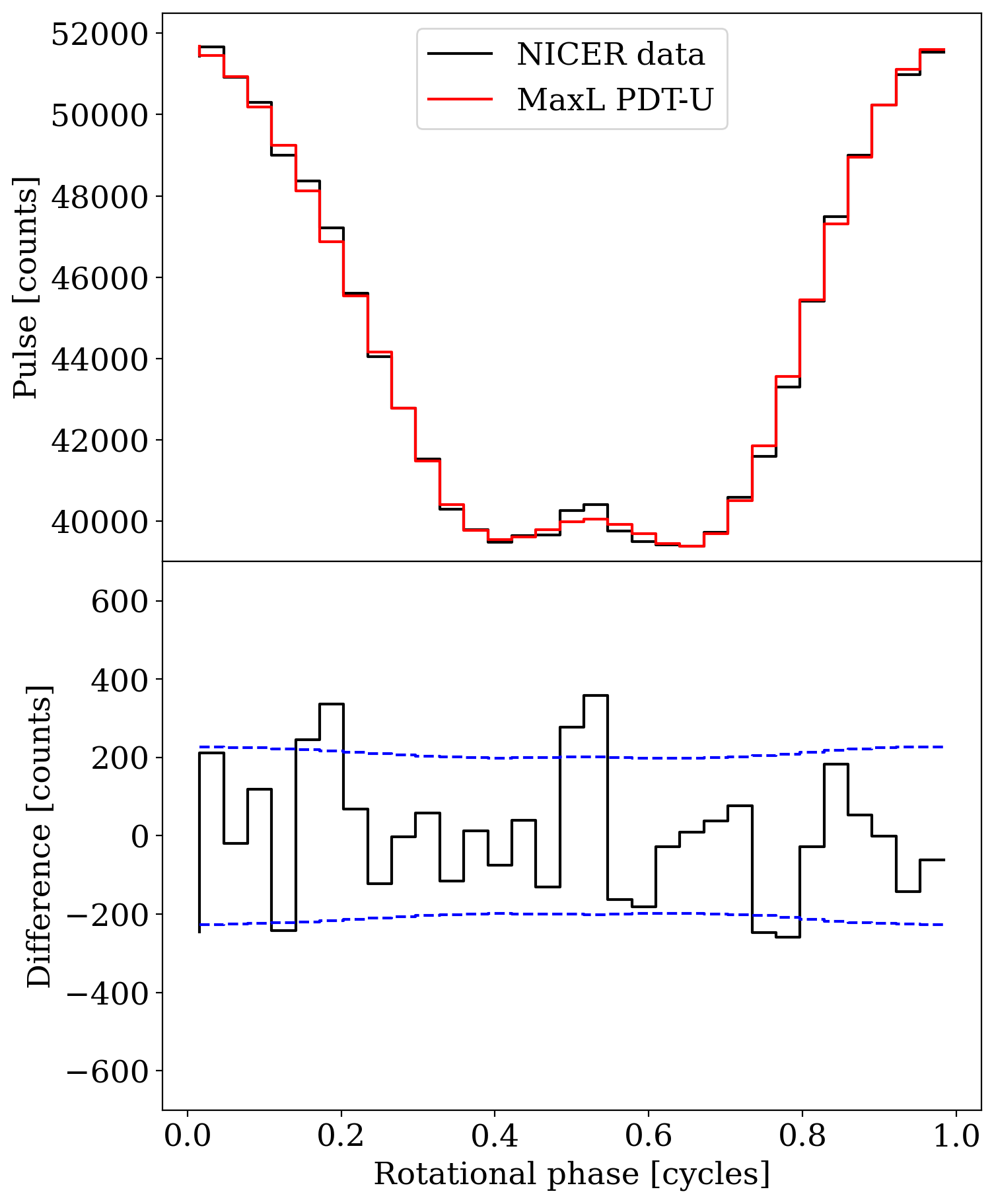}    
    \caption{\small{
    Bolometric pulse profiles (counts summed over energy channels) for the maximum-likelihood parameters (red-stepped curves) of the \texttt{ST-U} H $T_{\mathrm{else}}$ (upper left-hand panel) and \texttt{PDT-U} H $R_{\mathrm{eq}}\in [10,14]$ km (upper right-hand panel) km models. 
    See Table \ref{table:evidences_likelihoods} and Section \ref{sec:results_pdtu} for model explanations.
    The bolometric \NICER data are shown with black-stepped curves.
    The differences (data $-$ model) are shown in the bottom panels also with black-stepped curves.
    The expected Poisson fluctuation levels (for \NICER data) are shown there with blue-dashed curves. 
    We note that the model curves are rather similar between all the \texttt{ST-U} versions and between all the \texttt{PDT-U} versions.
    }}
    \label{fig:bolometric_pulses}
    \end{figure*}
}

{
    \begin{figure*}[!htbp]
    \centering
    \includegraphics[
    width=\textwidth]    
    {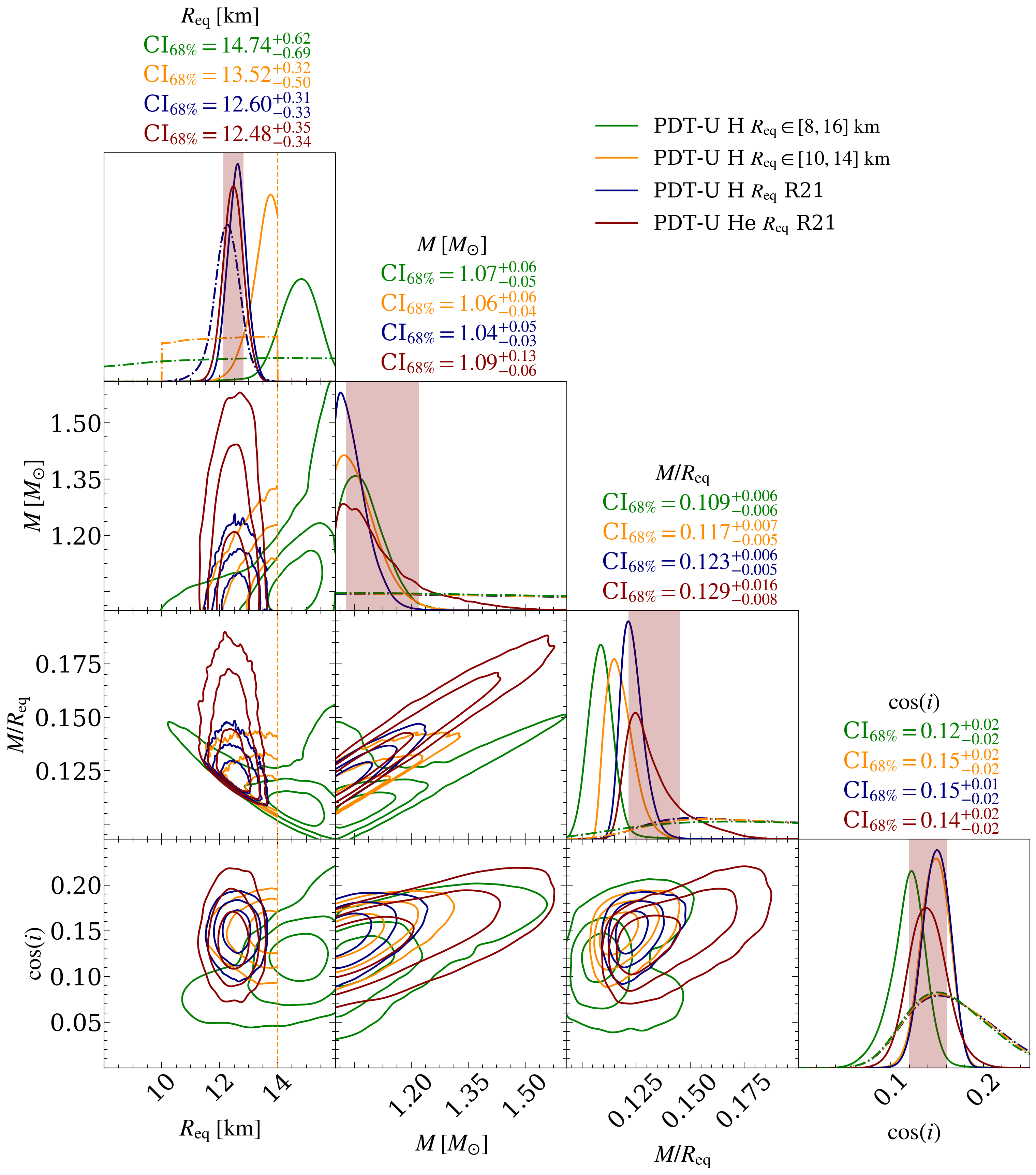} 
    \caption{\small{
    Radius, mass, compactness, and inclination posterior distributions using the \NICER and \xmm data sets for the \texttt{PDT-U} model.    
    The results with three different radius priors are shown, but
    the run with widest radius prior ($R_{\mathrm{eq}}\in [8,16]$ km) seems not to have converged because it finds only significantly worse fits to the data than the others.
    Dashed-dotted curves represent the marginal prior probability density functions (PDFs). 
    The vertical dashed orange line shows $R_{\mathrm{eq}}=14$ km to guide the eye. 
    The shaded vertical bands show the $68.3\%$ credible intervals (for the posteriors corresponding to the red curves), and the contours in the off-diagonal panels show the $68.3\%$, $95.4\%$, and $99.7\%$ credible regions. 
    See the captions of Figure 5 of \citet{Riley2021} for additional details about the figure elements.
    }}
    \label{fig:posteriors_spacetime}
    \end{figure*}
}

{
    \begin{figure}[t!]
    \centering
    \resizebox{\hsize}{!}{\includegraphics[
    width=\textwidth]{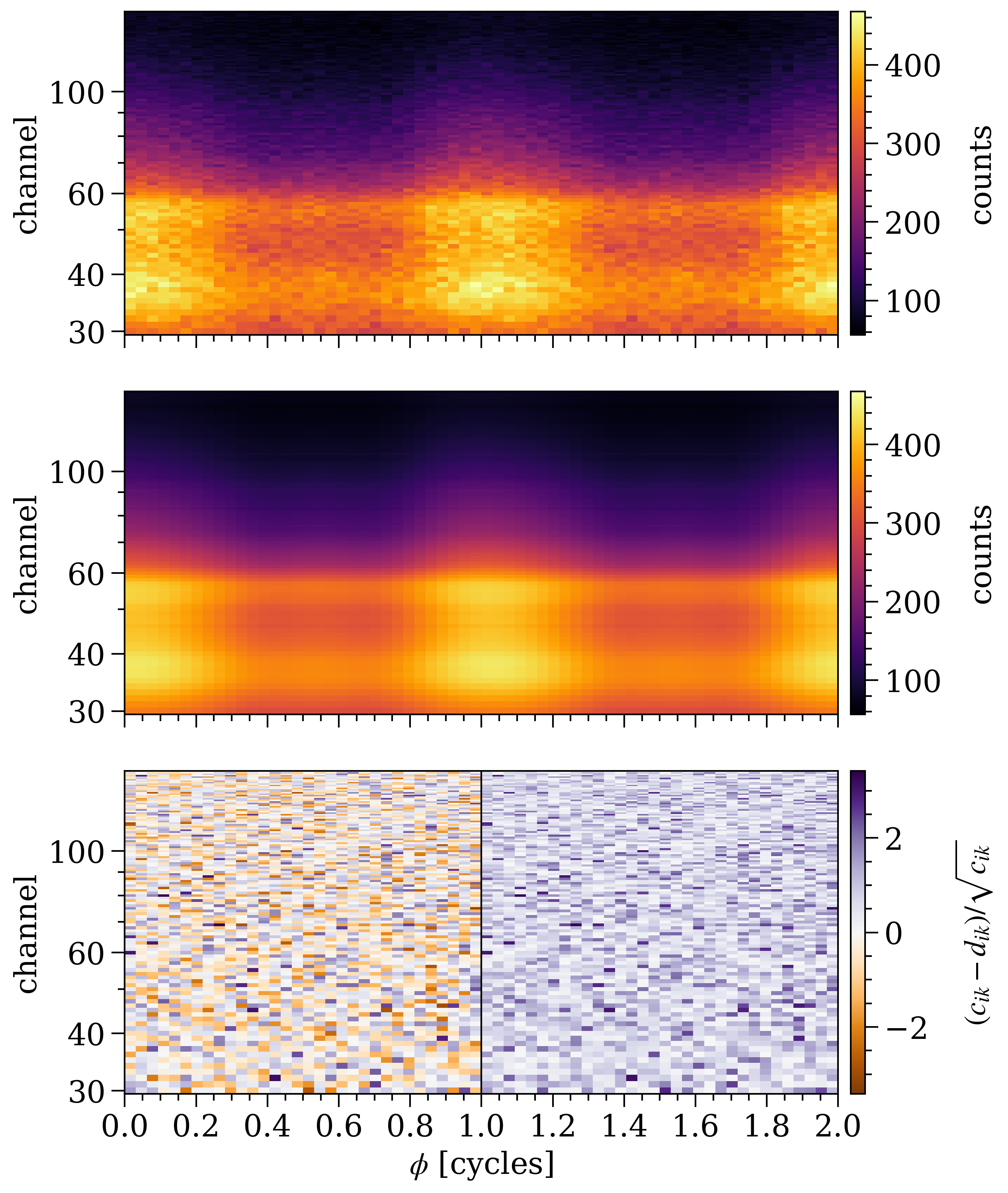}}
    \caption{\small{
The \NICER count data (top panel), posterior-expected count numbers (middle panel, averaged from 200 equally weighted posterior samples), and (Poisson) residuals (bottom panel) for the \texttt{PDT-U} H $R_{\mathrm{eq}}\in [10,14]$ km model. 
See Figure 6 of \citet{Riley2021} for additional details about the figure elements.
The complete figure set (seven images), which includes the residuals for all of the models listed in Table \ref{table:evidences_likelihoods}, is available in the online journal (HTML version). 
    }}
    \label{fig:residuals}
    \end{figure}
}

\begin{deluxetable}{lccc}[b]
\tablecaption{Model Performance Measures}
\tablehead{\multicolumn{1}{l}{Model} & \multicolumn{1}{l}{$\Delta\widehat{\ln\mathcal{Z}}$}  & \multicolumn{1}{l}{$\Delta\widehat{\ln\textrm{ML}}$} & \multicolumn{1}{l}{$\chi^{2}_{\textrm{ML,bol}}$}}
\startdata
\texttt{PDT-U} He $R_{\mathrm{eq}}$ R21 &
$0$ &
$0$ &
$18.2$ \\
\texttt{PDT-U H} $R_{\mathrm{eq}} \in [10, 14]$ km&
$-7.35$ &
$-2.39$ &
$19.8$ \\
\texttt{PDT-U H} $R_{\mathrm{eq}}$ R21 &
$-8.13$ &
$-2.72$ &
$19.5$ \\
\texttt{PDT-U H} $R_{\mathrm{eq}} \in [8, 16]$ km&
$-24.73$ &
$-10.16$ &
$21.7$ \\
\texttt{ST-U H} $T_{\mathrm{else}}$ &
$-32.56$ &
$-48.05$ &
$77.4$ \\
\texttt{ST-U} H $3\alpha$ &
$-35.98$ &
$-58.78$ &
$80.7$ \\
\texttt{ST-U} H &
$-36.56$ &
$-58.83$ &
$85.5$ \\
\enddata
\tablecomments{\ \ 
The evidence ($\widehat{\ln\mathcal{Z}}$) and maximum-likelihood ($\widehat{\ln\textrm{ML}}$) differences are shown compared to the \texttt{PDT-U} He $R_{\mathrm{eq}}$ R21 model. 
We also show the bolometric \NICER $\chi^{2}$ values ($\chi^{2}_{\textrm{ML,bol}}$) corresponding the maximum-likelihood sample.
Note that the exact d.o.f. is not known, but it is something smaller than $32$ (see Footnote \ref{footnote:dof}).
For \texttt{ST-U}, only runs with the broadest radius prior are considered here;  besides the standard case (\texttt{ST-U} H), values are also shown when including emission from rest of the star surface (\texttt{ST-U H} $T_{\mathrm{else}}$) and when having three different \xmm effective-area scaling factors (\texttt{ST-U} H $3\alpha$).
}
\end{deluxetable}\label{table:evidences_likelihoods}

{
    \begin{figure}[t!]
    \centering
    \resizebox{\hsize}{!}{\includegraphics[
    width=\textwidth]{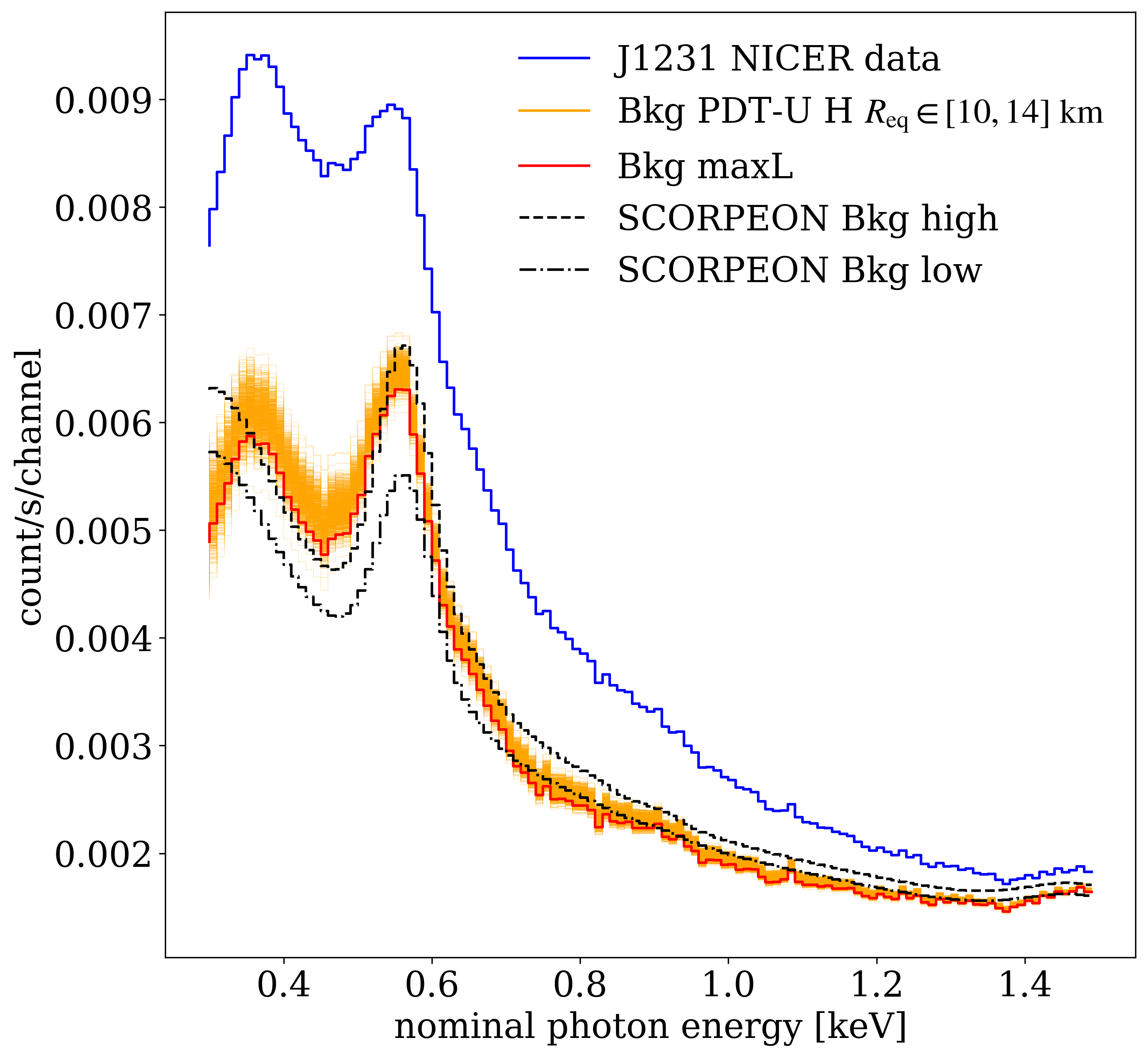}}
    \caption{\small{
Inferred \NICER background for the joint \NICER and \xmm analysis using the \texttt{PDT-U} H $R_{\mathrm{eq}}\in [10,14]$ km model.
The blue solid stepped curve shows the total \NICER count-rate spectrum.
The orange stepped curves show the backgrounds that maximize the likelihood for $1000$ equally weighted posterior samples.
The red curve shows the corresponding background for the maximum-likelihood sample. 
The black-dashed (dotted-dashed) curve shows the highest (lowest) background when considering the model uncertainty, as explained in Section \ref{sec:data_nicer}.
The inferred backgrounds of the other presented models are very similar to those shown here.
    }}
    \label{fig:bkg_inferred}
    \end{figure}
}

{
    \begin{figure}[t!]
    \centering
    \resizebox{\hsize}{!}{\includegraphics[
    width=\textwidth]{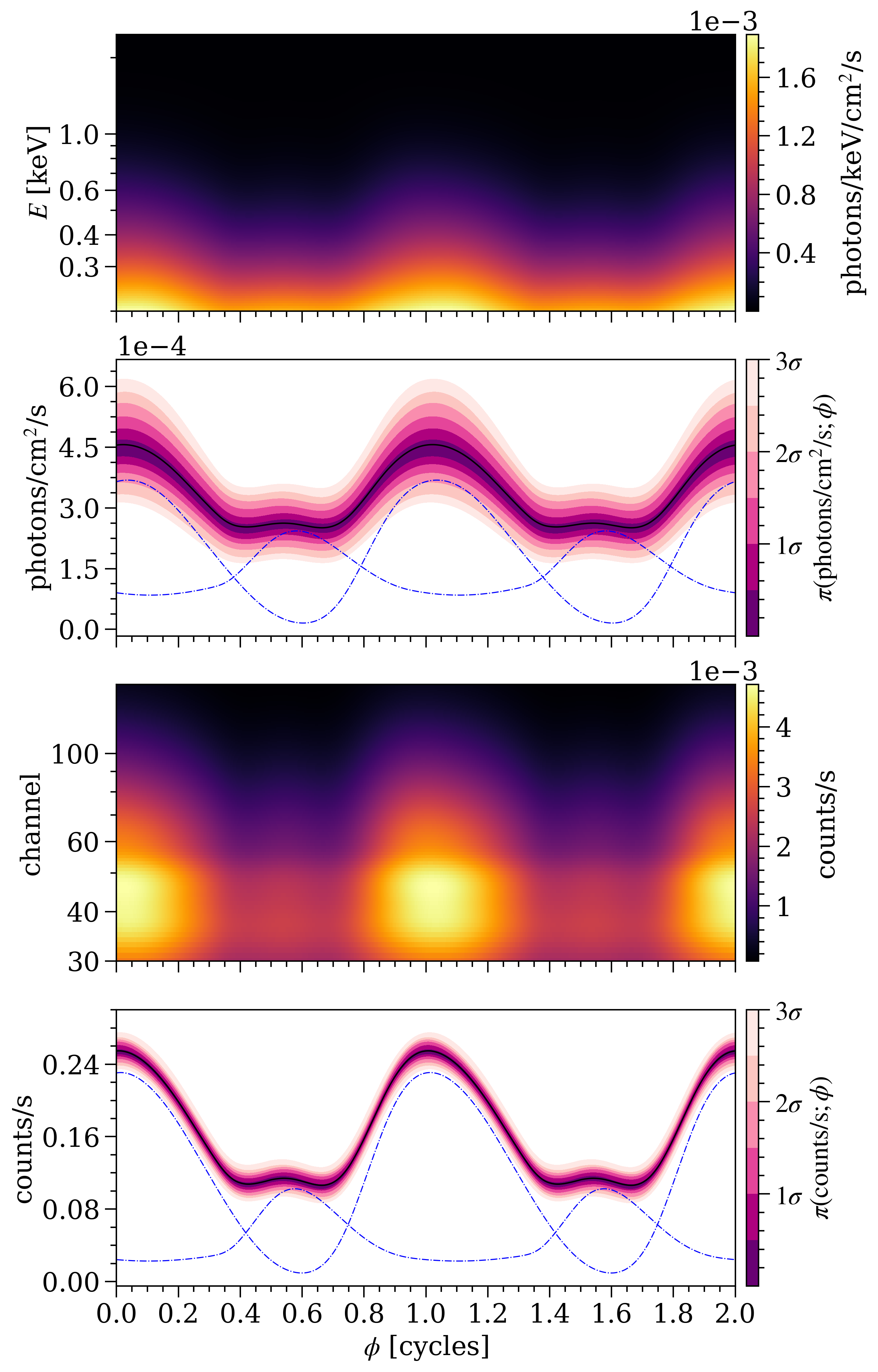}}
    \caption{\small{
Posterior-expected pulse profiles for the joint \NICER and \xmm analysis using the \texttt{PDT-U} H $R_{\mathrm{eq}}\in [10,14]$ km model.
The top and top-center panels show the signal incident on the telescopes, and the bottom-center and bottom panels show the signal registered by \NICER after accounting for the instrument response. 
The blue-dashed curves show the posterior-expected signals generated by the two hot regions separately and the black-solid curves in combination.
The conditional posterior distribution of the incident photon flux and the \NICER count rate are also presented (purple contours in the top-center and bottom panels, respectively).
See Figure 11 of \citet{Riley2021} for additional details about the figure elements.
    }}
    \label{fig:pulse_inferred}
    \end{figure}
}

{
    \begin{figure*}[t!]
    \centering
    \resizebox{\hsize}{!}{\includegraphics[
    width=\textwidth]{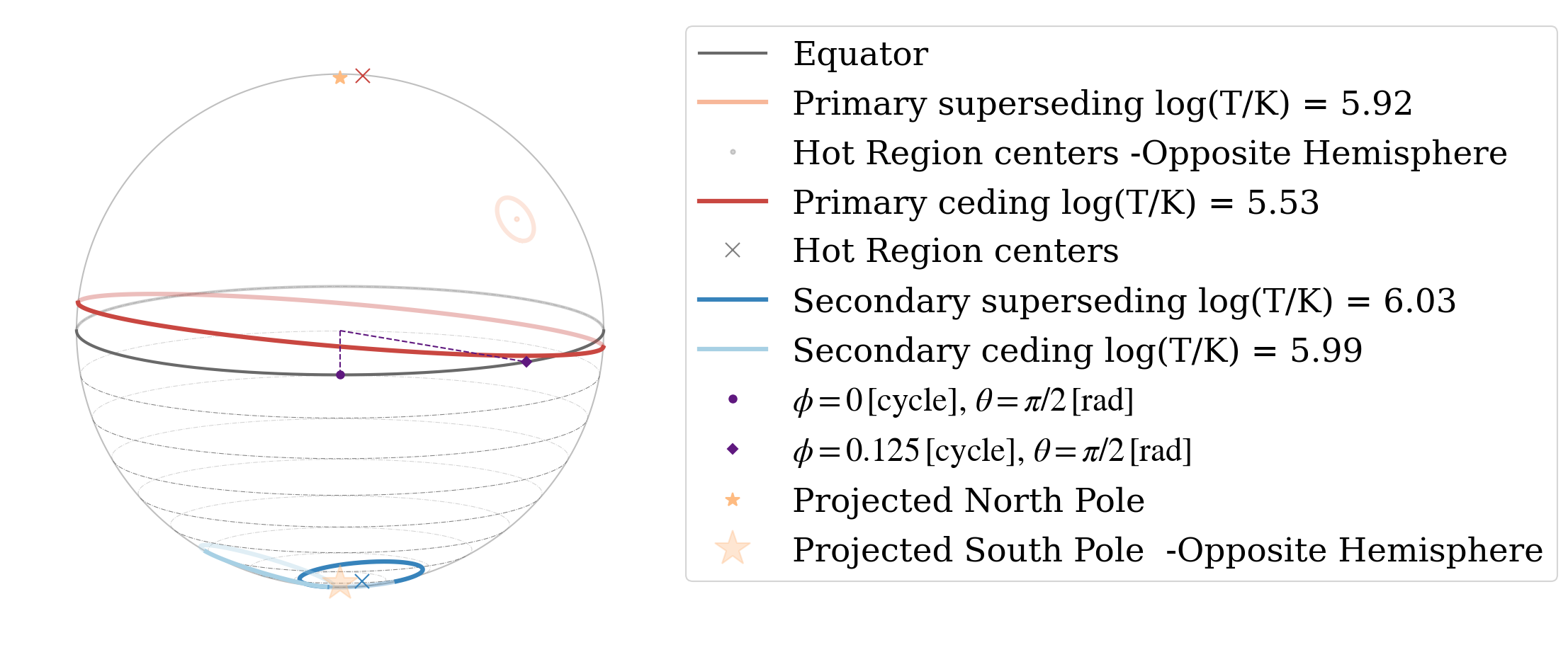}}
    \caption{\small{
The maximum-likelihood geometry configuration for the joint \NICER and \xmm analysis using the \texttt{PDT-U} H $R_{\mathrm{eq}}\in [10,14]$ km model.  
The viewing angle represents the Earth's inclination to the spin axis.
The maximum-likelihood geometry is very similar for the \texttt{PDT-U} H $R_{\mathrm{eq}}$ R21 and \texttt{PDT-U} He $R_{\mathrm{eq}}$ R21 models, except that the primary superseding component is located at the equator for the latter.
    }}
    \label{fig:geom_inferred}
    \end{figure*}
}

\section{Inferences}\label{sec:results}

\subsection{ST-U}\label{sec:results_stu}

We start with the inference for \joa using the \texttt{ST-U} model. 
We applied the model using three different versions of it: (1) Basic \texttt{ST-U}, (2) \texttt{ST-U} where the rest of the star's surface is also emitting (as mentioned in Section \ref{sec:pp_modeling}), (3) \texttt{ST-U} where all three \xmm instruments have their effective-area scaling factors as free parameters (as mentioned in Section \ref{sec:response}). 
The corresponding posteriors for the spacetime parameters are shown in the Appendix \ref{appendix:additional_figures} for the standard \XPSI radius prior (posteriors for the other parameters are available in the Zenodo, \citealt{salmi_zenodo24hotel} (doi:\href{https:/doi.org/10.5281/zenodo.13358349}{10.5281/zenodo.13358349})).
We found that in all three cases the inferred radius is very small, around $5$ km, and the NS is extremely compact, even though the mass is only slightly above $1$ \msol.
In fact, all these solutions strongly hit our prior upper bound on surface gravity set by the hydrogen atmosphere table limits ($\log_{10}g = 15.0$).

While no very obvious patterns were observed in the residuals between the inferred phase-energy resolved signals and the \joa data (which would indicate a poor fit to the data), we found that the fit to the phase-averaged (or bolometric) \NICER pulse profile is rather bad for all the \texttt{ST-U} models (see the left-hand panel of Figure \ref{fig:bolometric_pulses} for an example).
The bolometric $\chi^2$ values for maximum-likelihood solutions are around $80$ for $32$ phase bins.\footnote{The degree of freedom (d.o.f.) is not obvious, but it should be a few smaller than $32$ depending on how many of the model parameters influence the bolometric pulse profile \citep{Dittmann2024}. 
We also note that better bolometric $\chi^2$ values are found when looking at the $1000$ highest likelihood samples in each \texttt{ST-U} run because the best bolometric \NICER $\chi^2$ value does not exactly coincide with the maximum joint (2D) \NICER and \xmm likelihood sample. However, these $\chi^2$ values are still always above $70$.\label{footnote:dof}}
In addition, if forcing the radius to be above $8$ km (making a full run with the second radius prior case mentioned in Section \ref{sec:radius_prior}), both the maximum-likelihood and the corresponding bolometric $\chi^2$ values became significantly worse.
The former dropped by more than 20 in $\ln$-units and the latter increased to $\sim 160$ (results of this run are available in the Zenodo, \citealt{salmi_zenodo24hotel}).
Most notably, the best-fit model was no longer able to produce the interpulse around phase $0.5$ that is seen in the data (see Figure \ref{fig:data_hotel}).
This was easier for the high compactness star, due to multiple images of the hot spots contributing to the interpulse.
The highest Bayesian evidence for a \texttt{ST-U} model (the one including rest of the star emission) was also found to be significantly smaller than for any of the \texttt{PDT-U} models (by $8-33$ in $\ln$-units), so we therefore focus on the latter next.

\subsection{PDT-U}\label{sec:results_pdtu}

We present next the inference results for \joa using the \texttt{PDT-U} model. 
In Figure \ref{fig:posteriors_spacetime} we show the radius, mass, compactness, and $\cos i$ posteriors for three radius priors:
(1) using a lower limit of $8$ km (with the upper limit being $16$ km as usual); (2) limiting the radius between $10$ km and $14$ km, and (3) using an informative radius prior based on the results of \citet[][the R21 case from Section \ref{sec:radius_prior}]{Raaijmakers2021}.
The last was performed using both hydrogen (H) and helium (He) atmospheres (He R21 with 10,000 live points instead of the 20,000 used otherwise, but the H R21 results were checked and found to be close to identical with both settings).
In all cases, we report the $68.3\,\%$ credible intervals around the median.
For the $R_{\mathrm{eq}}\in [8,16]$ km case, we obtain $R_{\mathrm{eq}} = 14.74_{-0.69}^{+0.62}$ km and $M = 1.07_{-0.05}^{+0.06}$ \msol.
For the $R_{\mathrm{eq}}\in [10,14]$ km case, we obtain $R_{\mathrm{eq}} = 13.52_{-0.50}^{+0.32}$ km and $M = 1.06_{-0.04}^{+0.06}$ \msol.
For the H R21 case, we obtain $R_{\mathrm{eq}} = 12.60_{-0.33}^{+0.31}$ km and $M = 1.04_{-0.03}^{+0.05}$ \msol.
Finally, for the He R21 case, we obtain $R_{\mathrm{eq}} = 12.48_{-0.34}^{+0.35}$ km and $M = 1.09_{-0.06}^{+0.13}$ \msol.

As mentioned also in Section \ref{sec:posterior_computation}, only the R21 and $R_{\mathrm{eq}}\in [10,14]$ km results are likely converged in terms of improving sampler settings.
Both of them infer practically the same geometry solution (see the remaining posteriors in Appendix \ref{appendix:additional_figures}) and yield almost equally good fit quality and evidence (see Table \ref{table:evidences_likelihoods}).
On the contrary, the maximum-likelihood and evidence values are significantly worse for the $R_{\mathrm{eq}}\in [8,16]$ km prior run: by $7$ and $17$, respectively (both in $\ln$-units) when comparing to the corresponding R21 run.
This indicates that the wider radius prior run did not explore the parameter space sufficiently, and thus inferred a completely different (wrong) geometry solution. 
This was also the case in our early stage 4000 live point run using the standard (broadest) radius prior for \texttt{PDT-U} (with some older settings).
In that case, both $5$ km and $15$ km radius modes were visibly detected in the posteriors,\footnote{The former was better in the likelihood but the latter in the evidence.} but both of them are significantly worse than the R21 and $R_{\mathrm{eq}}\in [10,14]$ km results in terms of likelihood and evidence.
From Table \ref{table:evidences_likelihoods}, we also see that the evidence and maximum likelihood are highest for the He R21 run: by $8$ and $3$ in $\ln$-units, respectively, compared to the H R21 run.
All of the credible intervals for the two atmosphere options are, however, very consistent with each other (with He giving often broader intervals).
 
The performance of the \texttt{PDT-U} model was found to be good, both in terms of the 2D \NICER residuals (see Figure \ref{fig:residuals}) and the bolometric $\chi^{2}$ value of the maximum-likelihood parameter vector (see the right-hand panel of Figure \ref{fig:bolometric_pulses}).
The $\chi^{2}$ value was around $18$ for the He R21 run, $19-20$ for the H R21 and $R_{\mathrm{eq}}\in [10,14]$ km runs, and $21$ for the nonconverged H $R_{\mathrm{eq}}\in [8,16]$ km run (all again with $32$ phase bins).
We also checked if the inferred \NICER background (based mostly on simultaneous \xmm fits) is consistent with the predicted SCORPEON background in Figure \ref{fig:bkg_inferred}.
We see that it generally is, except at the lowest energies (close to $0.3$ keV) where the inferred background is somewhat below, at $0.4-0.6$ keV where it is somewhat above, and at $1.0-1.4$ keV where it is slightly below the SCORPEON range.
Inferring a slightly higher background is not surprising because the SCORPEON estimate does not include any unpulsed and unmodeled emission from the NS nor any emission from nearby sources (see the \xmm image in Figure 8 of \citealt{bogdanov19a}).
In addition, as noted in Section \ref{sec:data_nicer}, the SCORPEON background estimate was found to depend on the chosen source spectral model (especially at the lowest energies), and none of the tried XSPEC models matches exactly the model used in \XPSI and its inferred parameters.
Therefore, we consider the agreement reasonable. 
We also note that in our early \NICER-only runs (without \xmm) the inferred background was found to be substantially too large (covering all the unpulsed signal).

The contribution from the two different hot regions to the total pulse can be viewed from Figure \ref{fig:pulse_inferred}, both for the incident signal and for that registered by \NICER (without background). 
We see that the pulse from the primary region has a smoother shape than that from the secondary region, which remains mostly flat between the pulse peaks.
The corresponding inferred hot region geometry is shown in Figure \ref{fig:geom_inferred} for the maximum-likelihood sample of the \texttt{PDT-U} H $R_{\mathrm{eq}}\in [10,14]$ km run.
The primary region is located on the northern hemisphere and consists of a rather small hot component surrounded by a very large cold component.
The secondary region is located close to the southern pole and consists of almost equally hot but barely connected components with similar sizes.
We note that the maximum-likelihood geometry is very similar for the \texttt{PDT-U} H $R_{\mathrm{eq}}$ R21 and \texttt{PDT-U} He $R_{\mathrm{eq}}$ R21 runs, except that the small hot component of the primary region is located at the equator in the latter case (which is still consistent with the H cases considering the width of the credible intervals).

The inferred values for the remaining parameters, e.g., for $D$ and $N_{\mathrm{H}}$, are found to be well consistent with expectations (see Section \ref{sec:implications}).
For all the models, we find the distance to be around $D = 0.65\pm0.05$ kpc.
Inferred column densities show slightly more scatter, but all of the converged \texttt{PDT-U} runs provide roughly $N_{\mathrm{H}} = 1.8 \pm 0.5 \times 10^{20} \mathrm{cm}^{-2}$.

\section{Discussion}\label{sec:discussion}

As shown in Section \ref{sec:results}, our results are rather inconclusive due to the sampling challenges encountered when exploring the \texttt{PDT-U} model parameters with a wide prior for NS radius. 
However, at least using a radius prior based on the existing observations and theory \citep{Raaijmakers2021}, or limiting the radius between $10$ and $14$ km, seems to lead to converged results and provides the best fits we have found to the \joa data.
Next, we discuss what these results imply (in Section \ref{sec:implications}), what factors may make analyzing \joa harder than the previous \NICER sources (in Section \ref{sec:comp_others}), and some caveats (in Section \ref{sec:caveats}).

\subsection{Implications}\label{sec:implications}

One of the more robust findings of this work is that the preferred hot region model is \texttt{PDT-U} rather than \texttt{ST-U}.
This means that assuming two circular uniformly emitting hot regions is not sufficient and may lead to wrong estimates of the NS parameters.
Assuming a NS radius consistent with the previous EOS analysis and using the \texttt{PDT-U} model, the inferred hot region geometry is also constrained to be rather nonantipodal (see Figure \ref{fig:geom_inferred}), which implies that purely centered dipolar magnetic field is unlikely for \joa (as for other \NICER RMPs analyzed so far).
The large cold emission component surrounding the primary hot spot may also be associated with emission from the rest of the star's surface, hinting that using a \texttt{ST+PDT} model (see \citealt{Vinciguerra2024bravo} for a definition) and allowing the rest of the star surface to emit might lead to similar results.
However, this is expected to speed-up the computation only mildly, and was thus not tried here.

We can also note a few interesting details about the inferred mass and radius.
We see from Figure \ref{fig:posteriors_spacetime} that all of the inferred masses are quite small and close to our $1$ \msol lower limit (this limit is motivated by supernova theory, see \citealt{muller2024}, and by the description of the early evolution of a NS, see \citealt{Strobel1999}).
These mass constraints are notably tighter than the mass prior which we used from radio timing measurements but they are still in the region where the prior is highest (see Figure \ref{fig:mcosi_prior}).
The radius constraints, on the other hand, do not provide as much information gain over the prior, especially in the R21 case. 
There the Kullback–Leibler (KL) divergence\footnote{Representing prior-to-posterior information gain.} is $0.55$, implying a nonzero information given by the data.
In the H $R_{\mathrm{eq}}\in [10,14]$ km case, the KL-divergence of radius is notably better ($1.40$) but still worse than that of mass ($2.55$).

Additionally, our results offer some insights into the NS atmosphere composition by finding better evidence and likelihoods for helium rather than hydrogen when using the \texttt{PDT-U} model and R21 radius prior (see Section \ref{sec:results_pdtu}, the same was, however, also detected in preliminary nonrestricted \texttt{ST-U} analyses).
However, the difference is not large enough to be conclusive considering that hydrogen composition is a priori more probable based on the rapid sinking of heavier elements via diffusive gravitational separation \citep{AI1980,Hameury1983,Brown2002,ZP2002}.  
A helium atmosphere could occur if all the hydrogen was converted to helium via diffusive nuclear burning \citep{Wijngaarden2019}, or if the star had only accreted from a completely hydrogen-depleted companion star \citep[see][and the references therein]{Bogdanov2021, Salmi2023}.
Nevertheless, there is evidence that the companion of \joa is a white dwarf with a hydrogen envelope \citep{Bassa2016}.

The $N_{\mathrm{H}}$ column density can be estimated via independent means and compared to the inferred values. 
Neutral hydrogen maps predict values $\sim 3.4\times 10^{20}\,\mathrm{cm}^{-2}$ \citep{HI4PI16}, but this should be considered an approximate upper limit because this represents the column density integrated to the edge of the Galaxy in that direction.\footnote{For this, we used the \texttt{HEASARC} $N_{\mathrm{H}}$ tool at \url{https://heasarc.gsfc.nasa.gov/cgi-bin/Tools/w3nh/w3nh.pl}.} 
Other estimates of $N_{\mathrm{H}}$, at the distance of the pulsar, can be obtained from 3D maps of infrared or optical extinction $A_{V}$ or reddening $E\left(B-V\right)$.  
At the position and distance of \joa, $\sim 0.6$~kpc, we find $\sim 3.5\times 10^{20}\,\mathrm{cm}^{-2}$ based on the 3D dust maps of \citet[][\footnote{Query tool is available at \url{http://argonaut.skymaps.info/}.} this 3D map shows little dependence of $E\left(B-V\right)$ on the distance beyond 0.2~kpc]{green19}. Alternatively, with the 3D dust maps of \citet{lallement22} and \citet{vergely22},\footnote{Data are available on Vizier at (\url{https://vizier.cds.unistra.fr}) or via an online tool at \url{https://explore-platform.eu/sdas/about/gtomo}.} we find $\sim 3.4\times 10^{20}\,\mathrm{cm}^{-2}$. Both results make use of the relation of \cite{foight16} between $N_{\mathrm{H}}$ and $A_{V}$.  Another estimate can exploit the relation between the DM (number density of free electrons in the line of sight) and $N_{\mathrm{H}}$ \citep{he13}. 
For \joa, ${\rm DM}=8.09\,\mathrm{pc}\,\mathrm{cm}^{-3}$, which gives $\sim 2.4\times 10^{20}\,\mathrm{cm}^{-2}$. 

The $N_{\mathrm{H}}$ values inferred in this work are globally consistent with these independent measurements. 
It is important to keep in mind that the measurements deduced above should be considered carefully because the employed relations may be uncertain, especially at low-absorption values because the sampling of objects is much smaller for low values of $N_{\mathrm{H}}$.  
In other words, these independent measurements mostly bring support that the chosen prior on $N_{\mathrm{H}}$ is broad enough and does not bias the inferred parameters.

Our inferred distance is also found to be well consistent with the preliminary parallax distance measurement (Cromartie et al. 2024, in preparation), both being around or above $600$ pc.
We note though that our distance upper limit ($700$ pc) might be too restrictive because our posteriors peak only slightly below it and the parallax distance uncertainties can also allow higher distances, especially when accounting for the Lutz-Kelker bias \citep{LutzKulker1973,Verbiest2012}.
Using a more sophisticated distance prior is, however, left to future work.

In one of the \texttt{ST-U} runs we also explored the effect of separate effective-area scaling factors for all the \xmm instruments.
We found $\alpha_{\mathrm{pn}} = 0.93 \pm 0.09$, $\alpha_{\mathrm{MOS1}} = 0.98 \pm 0.09$, and $\alpha_{\mathrm{MOS2}} = 1.00 \pm 0.09$.
These are similar enough with each other that using one single $\alpha_{\mathrm{XMM}}$ seems justified for the final \texttt{PDT-U} models.

\subsection{Comparison to Other Pulsars}\label{sec:comp_others}

Compared to the other \NICER RMPs analyzed so far, constraining the NS parameters for \joa appears more challenging. There are several possible reasons for this. 
First of all, the available priors on mass and observer inclination are much less restrictive for \joa than for \joh and \jof.
For the isolated pulsar \jdbl, there is even less prior information about mass and inclination, but the distance is known much more precisely based on radio timing parallax measurements \citep{Arzoumanian2018}.
These factors, together with the relatively large number of free hot region geometry parameters needed for the \texttt{PDT-U} model, likely make the sampling problem more difficult. 
In the previous \XPSI analyses, the simple \texttt{ST-U} model was found to be adequate for \joh \citep{Riley2021} and the \texttt{CST+PDT} model, with three parameters less than for \texttt{PDT-U}, for \jof \citep{Choudhury24}.
In \citet{Vinciguerra2024bravo} \texttt{PDT-U} was one of the models applied to \jdbl, however its convergence was not investigated.

\joa also differs from the other pulsars also in terms of the pulse shape. 
As seen from Figures \ref{fig:data_hotel} and \ref{fig:pulse_inferred}, there is only a weak interpulse appearing in the bolometric pulse profile.
This feature is especially weak in the 2D data leading possibly to a multitude of solutions that can more or less fit it but that are too distinct in the parameter space to be explored thoroughly if one does not shrink the prior space.
For \jdbl and \joh, the secondary pulse is much stronger, and for \jof there is no peaky interpulse (see \citealt{bogdanov19a} and \citealt{Wolff21}). However, it could also be that the source properties of \joa simply differ from the others in such a way that they are more difficult to recover.  

It is nevertheless encouraging that despite the uncertainty, we are able to obtain solutions for this source that fit the data well and are also consistent with the results from other NICER RMPs and multimessenger analyses. 
There are clearly various steps that might improve the situation:  tighter priors, additional computational resources, or a longer data set with more photons. 
However, before devoting additional resources (in the form of observing or computational time) to this source, we would recommend a program of simulations using synthetic data to determine whether this is likely to lead to improvements or whether there are fundamental issues with this type of pulse profile that are likely to prevent efforts to obtain tighter and more robust constraints.

\subsection{Caveats}\label{sec:caveats}

The caveats of this work mostly follow those present in the previous \NICER analyses as well \citep[see e.g.,][for discussion]{Bogdanov2021}.
Additionally, we are now dependent on more restrictive radius priors, e.g., based on the results of \citet{Raaijmakers2021}, because the radius could not be constrained fully independently in this work.
Therefore, the existence of better solutions far from the R21 prior peak, which is around $12$ km, or at least outside from the [10,14] km range, cannot be ruled out (even though the hard limits for the R21 case were still kept at $\sim 4.4$ and $16$ km).

We also note that our results rely on $M - \cos i$ priors from a preliminary radio timing inference (Cromartie et al. 2024, in preparation). 
In the early analyses without these priors, we still found mostly similar (small) masses but the inclination angles were found to be much smaller, with $\cos i$ peaking between $0.4$ and $0.9$ (depending on the model) instead of at around $0.15$ as predicted by the radio timing.
However, these analyses did not yet employ any restricted radius prior for the \texttt{PDT-U} model.
We also note that the inclination angle from radio timing refers to the angle between the line of sight and the orbital axis, whereas we use it as the angle between line of sight and NS spin axis.
However, previous accretion of matter is expected to have gradually aligned the spin and orbital axes of the system (see, e.g., \citealt{Bhattacharya_Heuvel_1991}), and any remaining tilt between them is likely to be insignificant compared to the width of the prior. 

We have not tested if setting the lower limit of mass below $1$ \msol would affect our results.
Recent spectral modeling of HESS J1731$-$347 has suggested that some NSs could have such low masses \citep{Doroshenko2022}.
However, this interpretation relies on several assumptions, including the distance associated with the star and the use of a uniform-temperature carbon atmosphere model, as discussed by \citet{Alford2023}.
More importantly, this model is a poor fit to longer, better quality \xmm data from 2014 of the same source, while a two-blackbody model with a mass of, e.g., $1.4$ \msol, provides a good fit to the same data \citep{Alford2023}.
Therefore, current data do not support a mass less than $1$ \msol for HESS J1731$-$347.

Finally, we note that the inferred temperatures of some of the hot regions are rather cold (notably below $10^6$ K), making the effects or partial ionization more relevant in the NS atmosphere, whereas we assume it to be fully ionized both for the hydrogen and helium case.
However, the effect is still expected to be minor because the signal is dominated by the hottest regions.
In addition, switching between hydrogen and helium was not found to significantly alter the results (for the R21 case), implying only minor dependence on the atmosphere choice.

\section{Conclusions}
\label{sec:conclusions}

We have jointly analyzed \NICER and \xmm data for \joa using X-ray pulse profile modeling.
We found that the simple \texttt{ST-U} model with two uniformly emitting spherical caps is not sufficient to explain the data, especially the bolometric pulse profile.
Instead, good fits both in bolometric and energy-resolved data and significantly larger Bayesian evidence were found using the \texttt{PDT-U} model, where both hot regions consist of two circular components that are allowed to have different temperatures.
Unfortunately, we did not manage to get converged results using a wide NS radius prior with this computationally expensive model.
However, forcing the radius to be consistent with previous multimessenger observations and nuclear theory, we found $R_{\mathrm{eq}} = 12.6 \pm 0.3$ km and $M = 1.04_{-0.03}^{+0.05}$ \msol and a substantial improvement in evidence and likelihoods.
When assuming helium (instead of hydrogen) composition for the NS atmosphere, we were also able reproduce the data well and obtained very similar parameter constraints.
Equally good fits, and very similar results, were also found for the hydrogen atmosphere when applying an uninformative prior limited between $10$ and $14$ km, except that the radius was inferred to be $R_{\mathrm{eq}} = 13.5_{-0.5}^{+0.3}$ km.
In all the best-fitting models we infer a quite similar nonantipodal hot region geometry, implying that a pure dipolar magnetic field is not likely for \joa.

\section*{Acknowledgments}
This work was supported in part by NASA through the \NICER mission and the Astrophysics Explorers Program. 
T.S., A.L.W., D.C., Y.K., and S.V. acknowledge support from ERC Consolidator Grant No.~865768 AEONS (PI: Watts).  
The use of the national computer facilities in this research was subsidized by NWO Domain Science.
Astrophysics research at the Naval Research Laboratory is supported by the NASA Astrophysics Explorer Program.
S.G. acknowledges the support of the CNES.
W.C.G.H. acknowledges support through grant 80NSSC23K0078 from NASA.
This research has made use of data products and software provided by the High Energy Astrophysics Science Archive Research Center (HEASARC), which is a service of the Astrophysics Science Division at NASA/GSFC and the High Energy Astrophysics Division of the Smithsonian Astrophysical Observatory.
We thank Thomas Riley for help in the early stages of this project and Jason Farquhar for discussions about statistical fits to weak features.

\facilities{\NICER, XMM}

\software{Cython \citep{Behnel2011}, fgivenx \citep{Handley2018}, GetDist \citep{Lewis2019}, GNU Scientific Library (GSL; \citealt{Galassi2009}), HEASoft \citep{heasoft2014}, Matplotlib \citep{Hunter2007}, MPI for Python \citep{Dalcin2008}, \MultiNest \citep{multinest09}, nestcheck \citep{Higson2018JOSS}, NumPy \citep{Walt2011},\PyMultiNest \citep{PyMultiNest}, Python/C language \citep{Oliphant2007}, SciPy \citep{Jones}, \XPSI \citep{xpsi}.}

\bibliographystyle{aasjournal}
\bibliography{allbib}

\clearpage 
\appendix
\section{Additional Tables and Figures}\label{appendix:additional_figures}

We summarize the results for the best-fitting hydrogen models (\texttt{PDT-U} H $R_{\mathrm{eq}}\in [10,14]$ km and \texttt{PDT-U} H $R_{\mathrm{eq}}$ R21) in Tables \ref{table:pdtu_H_R1014} and  \ref{table:pdtu_H_R21}.
Additional \texttt{PDT-U} posteriors, including the parameters not shown in Figure \ref{fig:posteriors_spacetime}, are presented in Figures \ref{fig:posteriors_hot_regions1}, \ref{fig:posteriors_hot_regions2},  and \ref{fig:posteriors_others}.
The main \texttt{ST-U} posterior distributions for radius, compactness, and mass are shown in Figure \ref{fig:posteriors_stu_spacetime}.

\begin{deluxetable*}{lccccc}
\caption{Summary Table for the \texttt{PDT-U} H $R_{\mathrm{eq}}\in [10,14]$ km Run Described in Section \ref{sec:results_pdtu}}\label{table:pdtu_H_R1014}
\tablehead{
\colhead{Parameter} & \colhead{Description} & \colhead{Prior PDF (Density and Support)} & \colhead{$\widehat{\textrm{CI}}_{68\%}$} & \colhead{$\widehat{D}_{\textrm{KL}}$} & \colhead{$\widehat{\textrm{ML}}$}
}
\startdata
$P$ $[$ms$]$ &
Coordinate spin period &
$P=3.6839$, fixed &
$-$ &
$-$ &
$-$\\
\hline
$M$ $[M_{\odot}]$ &
gravitational mass &
$M \sim N(1.00233782,0.93010502)^{\mathrm{a}}$ &
$1.06_{-0.04}^{+0.06}$ &
$2.55$ &
$1.069$ \\ 
$\cos(i)$ &
Cosine Earth inclination to spin axis &
$\cos(i) \sim N(\mu_{\cos i}(M),\sigma_{\cos i} (M))^{\mathrm{b}} $ &
$0.15_{-0.02}^{+0.01}$ &
$0.95$ &
$0.172$ \\ 
\hline
$R_{\textrm{eq}}$ $[$km$]$ &
Coordinate equatorial radius &
$R_{\textrm{eq}}\sim U(10,14)$ &
$13.52_{-0.50}^{+0.32}$ &
$1.40$ &
$13.007$ \\ 
&With compactness condition & $R_{\textrm{polar}}/r_{\rm g}(M)>3$\\
&With surface gravity condition & $13.7\leq \log_{10}g(\theta)\leq15.0$,~$\forall\theta$\\
\hline
$\Theta_{p}$ $[$radians$]$ &
$p$ superseding component center colatitude &
$\cos(\Theta_{p})\sim U(-1,1)$ &
$1.29_{-0.21}^{+0.21}$ &
$0.93$ &
$1.239$ \\ 
$\Theta_{c,p}$ $[$radians$]$ &
$p$ ceding component center colatitude &
$\cos(\Theta_{c,p})\sim U(-1,1)$ &
$0.23_{-0.09}^{+0.13}$ &
$3.32$ &
$0.123$ \\ 
$\Theta_{s}$ $[$radians$]$ &
$s$ superseding component center colatitude &
$\cos(\Theta_{s})\sim U(-1,1)$ &
$2.78_{-0.03}^{+0.03}$ &
$4.72$ &
$2.761$ \\ 
$\Theta_{c,s}$ $[$radians$]$ &
$s$ ceding component center colatitude &
$\cos(\Theta_{c,s})\sim U(-1,1)$ &
$2.81_{-0.04}^{+0.04}$ &
$4.46$ &
$2.844$ \\ 
$\phi_{p}$ $[$cycles$]$ &
$p$ superseding component initial phase &
$\phi_{p}\sim U(-0.25,0.75)$, wrapped &
$0.38_{-0.01}^{+0.01}$ &
$5.25$ &
$0.375$\\ 
$\phi_{s}$ $[$cycles$]$ &
$s$ superseding component initial phase &
$\phi_{s}\sim U(-0.25,0.75)$, wrapped &
$0.53_{-0.01}^{+0.01}$ & 
$5.29$ &
$0.536$ \\ 
$\chi_{p}$ $[$radians$]$ &
Azimuthal offset between the $p$ components &
$\chi_{p}\sim U(-\pi,\pi)$ &
$-0.09_{-0.24}^{+0.26}$ &
$0.41$ &
$0.010$\\ 
$\chi_{s}$ $[$radians$]$ &
Azimuthal offset between the $s$ components &
$\chi_{s}\sim U(-\pi,\pi)$ &
$-1.63_{-0.05}^{+0.05}$ &
$6.89$ &
$-1.622$\\ 
$\zeta_{p}$ $[$radians$]$ &
$p$ superseding component angular radius &
$\zeta_{p}\sim U(0,\pi/2)$ &
$0.10_{-0.02}^{+0.02}$ &
$2.85$ &
$0.094$ \\ 
$\zeta_{c,p}$ $[$radians$]$ &
$p$ ceding component angular radius &
$\zeta_{c,p}\sim U(0,\pi/2)$ &
$1.44_{-0.15}^{+0.09}$ &
$4.28$ &
$1.548$ \\ 
$\zeta_{s}$ $[$radians$]$ &
$s$ superseding region angular radius &
$\zeta_{s}\sim U(0,\pi/2)$ &
$0.26_{-0.02}^{+0.03}$ &
$3.15$ &
$0.237$ \\ 
$\zeta_{c,s}$ $[$radians$]$ &
$s$ ceding component angular radius &
$\zeta_{s,p}\sim U(0,\pi/2)$ &
$0.28_{-0.02}^{+0.02}$ &
$3.16$ &
$0.266$ \\ 
&No region-exchange degeneracy & $\Theta_{s}\geq\Theta_{p}$\\
&Nonoverlapping hot regions & function of all $\Theta$, $\phi$, $\chi$, and $\zeta$\\
&Overlapping hot region components$^{\mathrm{c}}$ & function of all $\Theta$, $\phi$, $\chi$, and $\zeta$\\
\hline
$\log_{10}\left(T_{p}\;[\textrm{K}]\right)$ &
$p$ superseding component effective temperature  &
$\log_{10}\left(T_{p}\right)\sim U(5.1,6.8)$, \TT{NSX} limits &
$5.90_{-0.03}^{+0.02}$ &
$4.05$ &
$5.918$ \\ 
$\log_{10}\left(T_{c,p}\;[\textrm{K}]\right)$ &
$p$ ceding component effective temperature  &
$\log_{10}\left(T_{c,p}\right)\sim U(5.1,6.8)$, \TT{NSX} limits &
$5.53_{-0.03}^{+0.03}$ &
$3.82$ &
$5.531$ \\ 
$\log_{10}\left(T_{s}\;[\textrm{K}]\right)$ &
$s$ superseding component effective temperature  &
$\log_{10}\left(T_{s}\right)\sim U(5.1,6.8)$, \TT{NSX} limits &
$6.02_{-0.01}^{+0.01}$ &
$5.50$ &
$6.027$ \\ 
$\log_{10}\left(T_{c,s}\;[\textrm{K}]\right)$ &
$s$ ceding component effective temperature  &
$\log_{10}\left(T_{c,s}\right)\sim U(5.1,6.8)$, \TT{NSX} limits &
$5.96_{-0.02}^{+0.02}$ &
$4.54$ &
$5.987$ \\ 
$D$ $[$kpc$]$ &
Earth distance &
$D\sim U(0.1,0.7)$ &
$0.65_{-0.05}^{+0.04}$ &
$2.19$ &
$0.620$ \\ 
$N_{\textrm{H}}$ $[10^{20}$cm$^{-2}]$ &
Interstellar neutral H column density &
$N_{\textrm{H}}\sim U(0,10)$ &
$1.77_{-0.50}^{+0.50}$ &
$2.29$ &
$2.101$ \\ 
\hline
$\alpha_{\rm{NICER}}$ &
\NICER effective-area scaling &
$\alpha_{\rm{NICER}},\alpha_{\rm{XMM}}\sim N(\boldsymbol{\mu},\boldsymbol{\Sigma})$ &
$0.98_{-0.07}^{+0.09}$ &
$0.14$ &
$1.071$\\ 
$\alpha_{\rm{XMM}}$ &
\xmm effective-area scaling &
$\alpha_{\rm{NICER}},\alpha_{\rm{XMM}}\sim N(\boldsymbol{\mu},\boldsymbol{\Sigma})$ &
$0.96_{-0.08}^{+0.08}$ &
$0.22$ &
$0.970$ \\ 
&With joint prior PDF $N(\boldsymbol{\mu},\boldsymbol{\Sigma})$  & $\boldsymbol{\mu}=[1.0,1.0]^{\top}$\\
&&$\boldsymbol{\Sigma}=
\begin{bmatrix}
0.104^{2} & 0.100^{2} \\
0.100^{2} & 0.104^{2}
\end{bmatrix}$\\
\hline
\hline
&Sampling Process Information&&& \\
\hline
&Number of free parameters: $23$ &&& \\
&Number of processes (multimodes): $4$ &&& \\
&Number of live points: $2\times10^{4}$ &&& \\
&Sampling efficiency (SE): $0.1$ &&& \\
&Termination condition: $0.1$ &&& \\
&Evidence: $\widehat{\ln\mathcal{Z}}= -21352.42\pm0.56$ &&&\\ 
&Number of core hours: $5.34533\times10^{5}$ &&& \\ 
&Likelihood evaluations: $4.83928845\times10^{8}$ &&& \\
\enddata
\tablecomments{\ \
We show the prior PDFs, $68.3\,\%$ credible intervals around the median $\widehat{\textrm{CI}}_{68\%}$, KL-divergence $\widehat{D}_{\textrm{KL}}$ in \textit{bits}, and the maximum-likelihood nested sample $\widehat{\textrm{ML}}$ for all the parameters. 
The subscripts $p$ and $s$ denote for primary and secondary hot region parameters, respectively. Note that $\phi_p$ ($\phi_s$) is measured with respect to the meridian on which Earth (Earth antipode) lies.
\\
$^{\mathrm{a}}$ Truncated between $1$ and $3$ \msol. However, note that for each $\cos i$ the $M$ prior is different, as seen in Figure \ref{fig:mcosi_prior}. \\
$^{\mathrm{b}}$ See the definitions of $\mu_{\cos i}$ and $\sigma_{\cos i}$ in Equations \eqref{eq:mu_cosi} and \eqref{eq:sigma_cosi}. \\
$^{\mathrm{c}}$ Ensuring that superseding components never engulf the corresponding ceding components.
\\
}
\end{deluxetable*}

\begin{deluxetable*}{lccccc}
\caption{Summary Table for the \texttt{PDT-U} H $R_{\mathrm{eq}}$ R21 Run Described in Section \ref{sec:results_pdtu}}\label{table:pdtu_H_R21}
\tablehead{
\colhead{Parameter} & \colhead{Description} & \colhead{Prior PDF (Density and Support)} & \colhead{$\widehat{\textrm{CI}}_{68\%}$} & \colhead{$\widehat{D}_{\textrm{KL}}$} & \colhead{$\widehat{\textrm{ML}}$}
}
\startdata
$P$ $[$ms$]$ &
Coordinate spin period &
$P=3.6839$, fixed & 
$-$ &
$-$ &
$-$\\
\hline
$M$ $[M_{\odot}]$ &
Gravitational mass &
$M \sim N(1.00233782,0.93010502)^{\mathrm{a}}$ &
$1.04_{-0.03}^{+0.05}$ &
$3.00$ &
$1.002$ \\ 
$\cos(i)$ &
Cosine Earth inclination to spin axis &
$\cos(i) \sim N(\mu_{\cos i}(M),\sigma_{\cos i} (M))^{\mathrm{b}} $ &
$0.15_{-0.02}^{+0.01}$ &
$1.02$ &
$0.165$ \\ 
\hline
$R_{\textrm{eq}}$ $[$km$]$ &
Coordinate equatorial radius &
\citet{Raaijmakers2021} $^{\mathrm{c}}$ &
$12.60_{-0.33}^{+0.31}$ &
$0.55$ &
$12.859$ \\ 
&With compactness condition & $R_{\textrm{polar}}/r_{\rm g}(M)>3$\\
&With surface gravity condition & $13.7\leq \log_{10}g(\theta)\leq15.0$,~$\forall\theta$\\
\hline
$\Theta_{p}$ $[$radians$]$ &
$p$ superseding component center colatitude &
$\cos(\Theta_{p})\sim U(-1,1)$ &
$1.34_{-0.21}^{+0.19}$ &
$1.01$ &
$1.172$ \\ 
$\Theta_{c,p}$ $[$radians$]$ &
$p$ ceding component center colatitude &
$\cos(\Theta_{c,p})\sim U(-1,1)$ &
$0.26_{-0.10}^{+0.16}$ &
$2.90$ &
$0.155$ \\ 
$\Theta_{s}$ $[$radians$]$ &
$s$ superseding component center colatitude &
$\cos(\Theta_{s})\sim U(-1,1)$ &
$2.77_{-0.03}^{+0.03}$ &
$4.67$ &
$2.716$ \\ 
$\Theta_{c,s}$ $[$radians$]$ &
$s$ ceding component center colatitude &
$\cos(\Theta_{c,s})\sim U(-1,1)$ &
$2.79_{-0.04}^{+0.04}$ &
$4.35$ &
$2.827$ \\ 
$\phi_{p}$ $[$cycles$]$ &
$p$ superseding component initial phase &
$\phi_{p}\sim U(-0.25,0.75)$, wrapped &
$0.38_{-0.01}^{+0.01}$ &
$5.30$ &
$0.375$\\ 
$\phi_{s}$ $[$cycles$]$ &
$s$ superseding component initial phase &
$\phi_{s}\sim U(-0.25,0.75)$, wrapped &
$0.53_{-0.01}^{+0.01}$ & 
$5.31$ &
$0.541$ \\ 
$\chi_{p}$ $[$radians$]$ &
Azimuthal offset between the $p$ components &
$\chi_{p}\sim U(-\pi,\pi)$ &
$-0.04_{-0.26}^{+0.25}$ &
$0.37$ &
$0.219$\\ 
$\chi_{s}$ $[$radians$]$ &
Azimuthal offset between the $s$ components &
$\chi_{s}\sim U(-\pi,\pi)$ &
$-1.65_{-0.05}^{+0.05}$ &
$6.95$ &
$-1.550$\\ 
$\zeta_{p}$ $[$radians$]$ &
$p$ superseding component angular radius &
$\zeta_{p}\sim U(0,\pi/2)$ &
$0.12_{-0.02}^{+0.02}$ &
$2.82$ &
$0.093$ \\ 
$\zeta_{c,p}$ $[$radians$]$ &
$p$ ceding component angular radius &
$\zeta_{c,p}\sim U(0,\pi/2)$ &
$1.44_{-0.14}^{+0.09}$ &
$4.40$ &
$1.412$ \\ 
$\zeta_{s}$ $[$radians$]$ &
$s$ superseding region angular radius &
$\zeta_{s}\sim U(0,\pi/2)$ &
$0.27_{-0.02}^{+0.03}$ &
$3.20$ &
$0.234$ \\ 
$\zeta_{c,s}$ $[$radians$]$ &
$s$ ceding component angular radius &
$\zeta_{s,p}\sim U(0,\pi/2)$ &
$0.29_{-0.02}^{+0.02}$ &
$3.21$ &
$0.285$ \\ 
&No region-exchange degeneracy & $\Theta_{s}\geq\Theta_{p}$\\
&Nonoverlapping hot regions & function of all $\Theta$, $\phi$, $\chi$, and $\zeta$\\
&Overlapping hot region components$^{\mathrm{d}}$ & function of all $\Theta$, $\phi$, $\chi$, and $\zeta$\\
\hline
$\log_{10}\left(T_{p}\;[\textrm{K}]\right)$ &
$p$ superseding component effective temperature  &
$\log_{10}\left(T_{p}\right)\sim U(5.1,6.8)$, \TT{NSX} limits &
$5.90_{-0.02}^{+0.02}$ &
$4.09$ &
$5.920$ \\ 
$\log_{10}\left(T_{c,p}\;[\textrm{K}]\right)$ &
$p$ ceding component effective temperature  &
$\log_{10}\left(T_{c,p}\right)\sim U(5.1,6.8)$, \TT{NSX} limits &
$5.53_{-0.03}^{+0.02}$ &
$3.77$ &
$5.556$ \\ 
$\log_{10}\left(T_{s}\;[\textrm{K}]\right)$ &
$s$ superseding component effective temperature  &
$\log_{10}\left(T_{s}\right)\sim U(5.1,6.8)$, \TT{NSX} limits &
$6.02_{-0.01}^{+0.01}$ &
$5.55$ &
$6.018$ \\ 
$\log_{10}\left(T_{c,s}\;[\textrm{K}]\right)$ &
$s$ ceding component effective temperature  &
$\log_{10}\left(T_{c,s}\right)\sim U(5.1,6.8)$, \TT{NSX} limits &
$5.96_{-0.02}^{+0.02}$ &
$4.50$ &
$5.980$ \\ 
$D$ $[$kpc$]$ &
Earth distance &
$D\sim U(0.1,0.7)$ &
$0.65_{-0.05}^{+0.03}$ &
$2.41$ &
$0.633$ \\ 
$N_{\textrm{H}}$ $[10^{20}$cm$^{-2}]$ &
Interstellar neutral H column density &
$N_{\textrm{H}}\sim U(0,10)$ &
$1.62_{-0.48}^{+0.47}$ &
$2.35$ &
$1.905$ \\ 
\hline
$\alpha_{\rm{NICER}}$ &
\NICER effective-area scaling &
$\alpha_{\rm{NICER}},\alpha_{\rm{XMM}}\sim N(\boldsymbol{\mu},\boldsymbol{\Sigma})$ &
$0.98_{-0.08}^{+0.09}$ &
$0.13$ &
$1.040$\\ 
$\alpha_{\rm{XMM}}$ &
\xmm effective-area scaling &
$\alpha_{\rm{NICER}},\alpha_{\rm{XMM}}\sim N(\boldsymbol{\mu},\boldsymbol{\Sigma})$ &
$0.95_{-0.08}^{+0.08}$ &
$0.26$ &
$0.950$ \\ 
&With joint prior PDF $N(\boldsymbol{\mu},\boldsymbol{\Sigma})$  & $\boldsymbol{\mu}=[1.0,1.0]^{\top}$\\
&&$\boldsymbol{\Sigma}=
\begin{bmatrix}
0.104^{2} & 0.100^{2} \\
0.100^{2} & 0.104^{2}
\end{bmatrix}$\\
\hline
\hline
&Sampling Process Information&&& \\
\hline
&Number of free parameters: $23$ &&& \\
&Number of processes (multimodes): $4$ &&& \\
&Number of live points: $2\times10^{4}$ &&& \\
&Sampling efficiency (SE): $0.1$ &&& \\
&Termination condition: $0.1$ &&& \\
&Evidence: $\widehat{\ln\mathcal{Z}}= -21353.20\pm0.56$ &&&\\ 
&Number of core hours: $2.27230\times10^{5}$ &&& \\
&Likelihood evaluations: $5.47847609\times10^{8}$ &&& \\
\enddata
\tablecomments{\ \
We show the prior PDFs, $68.3\,\%$ credible intervals around the median $\widehat{\textrm{CI}}_{68\%}$, KL-divergence $\widehat{D}_{\textrm{KL}}$ in \textit{bits}, and the maximum-likelihood nested sample $\widehat{\textrm{ML}}$ for all the parameters. 
The subscripts $p$ and $s$ denote for primary and secondary hot region parameters, respectively. Note that $\phi_p$ ($\phi_s$) is measured with respect to the meridian on which Earth (Earth antipode) lies.
\\
$^{\mathrm{a}}$ Truncated between $1$ and $3$ \msol. However, note that for each $\cos i$ the $M$ prior is different, as seen in Figure \ref{fig:mcosi_prior}. \\
$^{\mathrm{b}}$ See the definitions of $\mu_{\cos i}$ and $\sigma_{\cos i}$ in Equations \eqref{eq:mu_cosi} and \eqref{eq:sigma_cosi}. \\
$^{\mathrm{c}}$ Truncated between $3r_{\rm g}(1) \approx 4.4$ km and $16$ km. The prior density corresponds to the R21 case that was explained in Section \ref{sec:radius_prior}.\\
$^{\mathrm{d}}$ Ensuring that superseding components never engulf the corresponding ceding components.
\\
}
\end{deluxetable*}

{
    \begin{figure*}[t!]
    \centering
    \includegraphics[
    width=\textwidth]
    {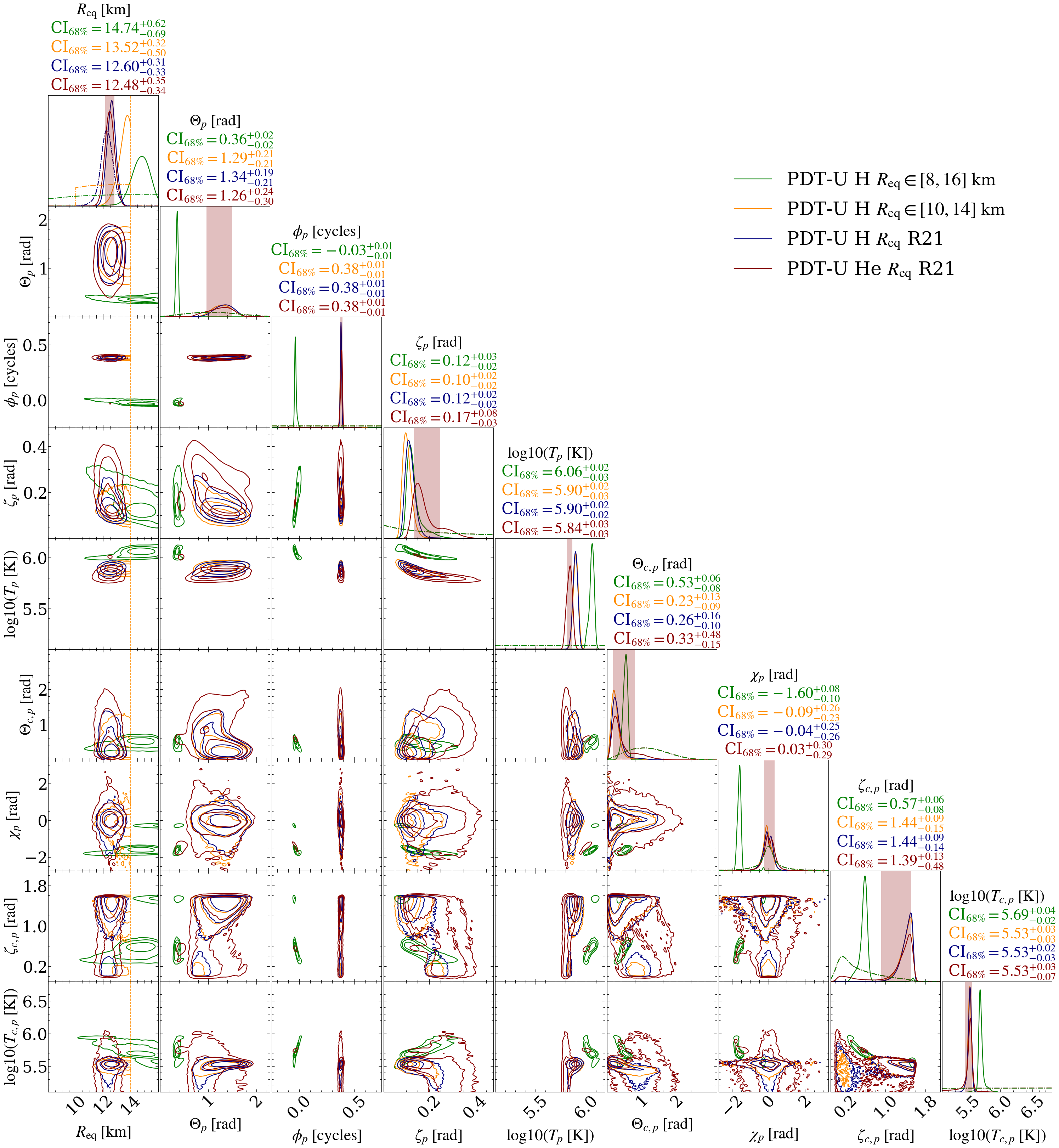}  
    \caption{\small{
    \texttt{PDT-U} posterior distributions for radius and the primary hot region parameters.
    See the caption of Figure \ref{fig:posteriors_spacetime} for additional details.    
    }}
    \label{fig:posteriors_hot_regions1}
    \end{figure*}
}

{
    \begin{figure*}[t!]
    \centering
    \includegraphics[
    width=\textwidth]
    {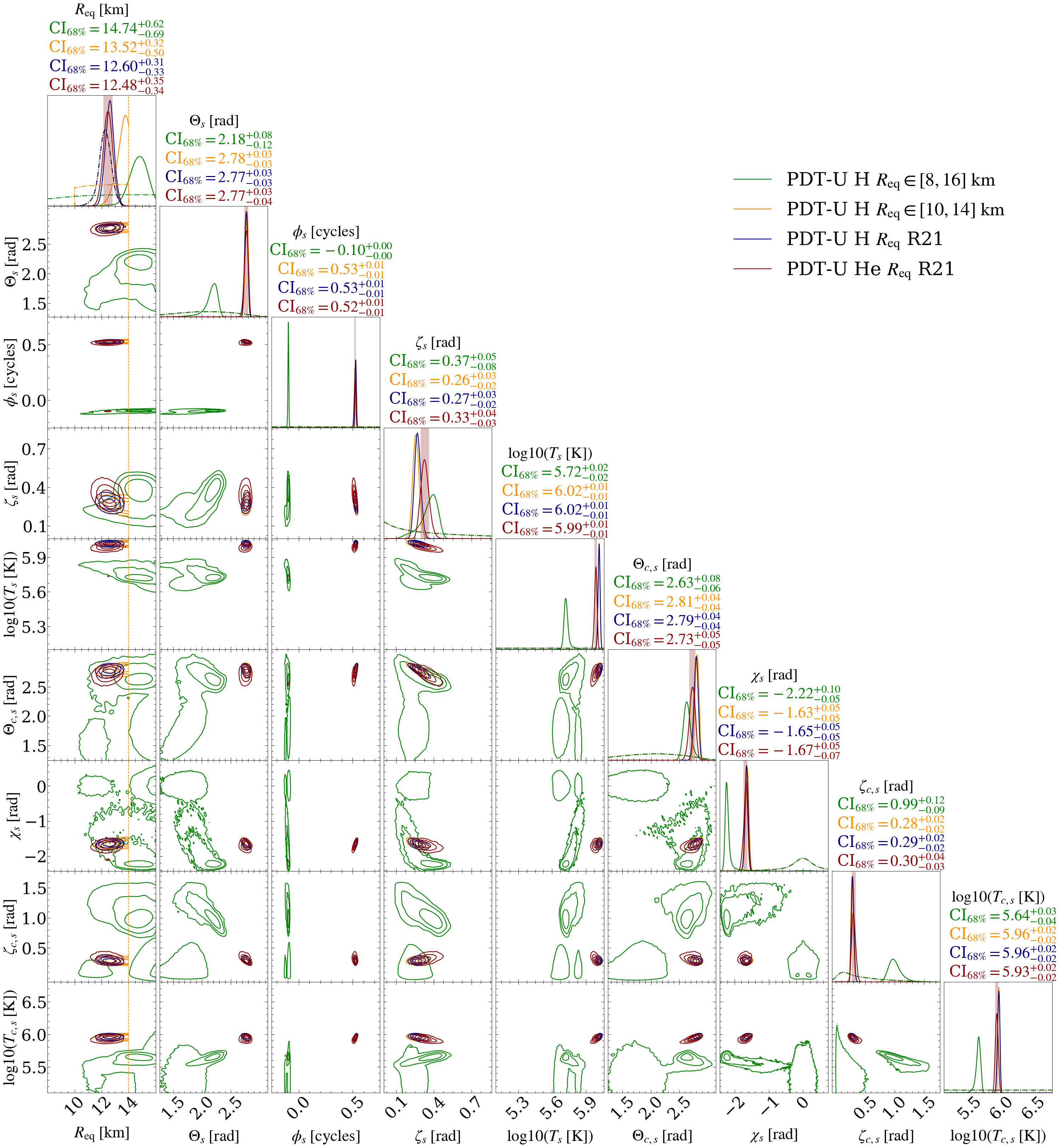}  
    \caption{\small{
    \texttt{PDT-U} posterior distributions for radius and the secondary hot region parameters.
    See the caption of Figure \ref{fig:posteriors_spacetime} for additional details.    
    }}
    \label{fig:posteriors_hot_regions2}
    \end{figure*}
}

{
    \begin{figure*}[t!]
    \centering
    \includegraphics[
    width=\textwidth]
    {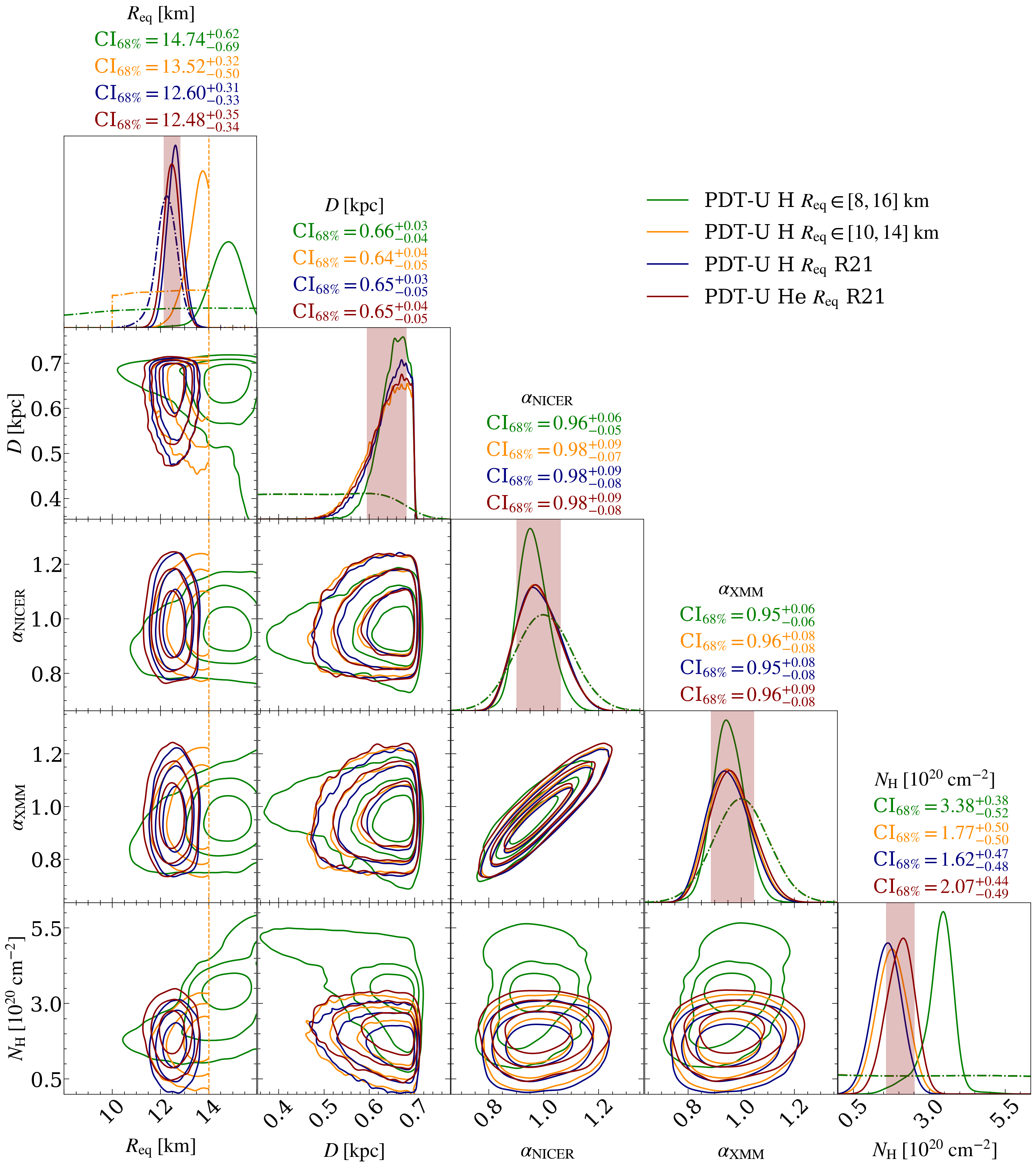}  
    \caption{\small{
    \texttt{PDT-U} posterior distributions for radius and the other parameters.
    See the caption of Figure \ref{fig:posteriors_spacetime} for additional details.    
    }}
    \label{fig:posteriors_others}
    \end{figure*}
}

{
    \begin{figure*}[t!]
    \centering
    \includegraphics[
    width=\textwidth]
    {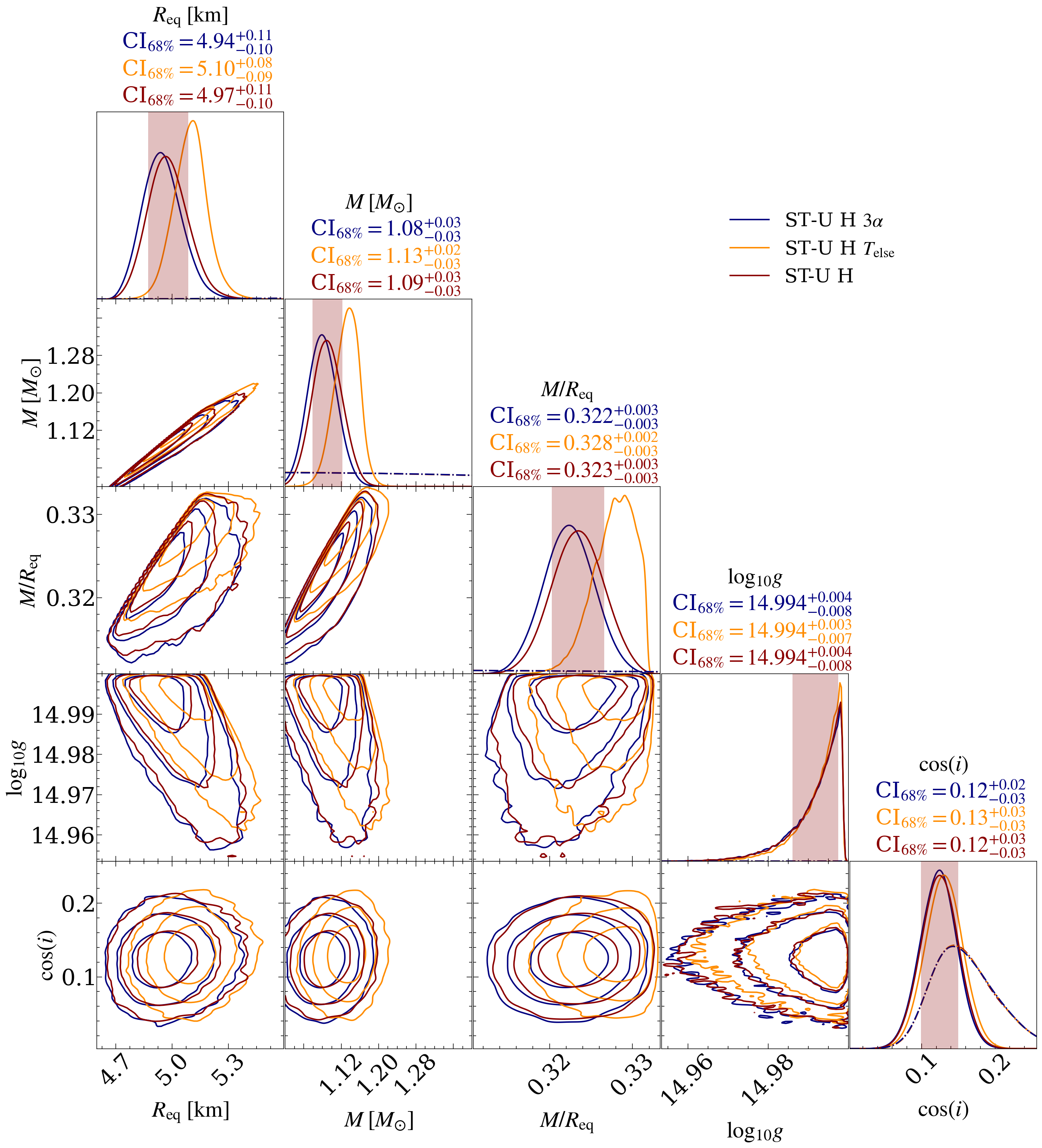}  
    \caption{\small{
    Posterior distributions for radius, mass, compactness, the logarithm of the surface gravity, and the cosine of the inclination for the \texttt{ST-U} runs.
    See the caption of Figure \ref{fig:posteriors_spacetime} for additional details.
    }}
    \label{fig:posteriors_stu_spacetime}
    \end{figure*}
}

\end{document}